\documentclass[iop,apj,twocolappendix]{emulateapj}

\pdfoutput=1

\usepackage{graphicx}
\usepackage{apjfonts}
\usepackage[UKenglish]{isodate}
\usepackage{multirow}
\usepackage[fleqn]{amsmath}
\usepackage{amssymb}
\usepackage{xfrac}

\usepackage[breaklinks,colorlinks,citecolor=blue,linkcolor=blue,urlcolor=blue,pdftex,bookmarks=true]{hyperref}
\usepackage[all]{hypcap}
\hypersetup{
    pdftitle={HSTPROMO GCs. IV. Kinematic profiles and average masses of blue straggler stars},
    pdfauthor={A.T. Baldwin, et al.}
}

\def\equationautorefname~#1\null{%
  equation~(#1)\null
}
\newcommand {\Msun} {M$_\odot$}
\newcommand {\afe} {\rm [\alpha/Fe]}
\newcommand {\feh} {\rm [Fe/H]}
\newcommand {\sigc} {\sigma_\mathrm{c}}

\newcommand {\Ri} {R_\mathrm{i}}
\newcommand {\sigi} {\sigma_\mathrm{i}}
\newcommand {\dsigi} {\Delta\sigma_\mathrm{i}}
\newcommand {\Li} {\mathcal{L}_\mathrm{i}}
\newcommand {\Posti} {\mathcal{P}_\mathrm{i}}

\newcommand {\Mbss} {M_\mathrm{BSS}}
\newcommand {\Mto} {M_\mathrm{MSTO}}
\newcommand {\sigbss} {\sigma_\mathrm{BSS}}
\newcommand {\sigto} {\sigma_\mathrm{MSTO}}

\newcommand {\nrel} {n_\mathrm{rel}}
\newcommand {\Tage} {T_\mathrm{age}}
\newcommand {\Trc} {T_\mathrm{rc}}
\newcommand {\Meq} {M_\mathrm{eq}}

\defcitealias{bellini2014}{Paper 1}
\defcitealias{watkins2015a}{Paper 2}

\begin{document}

\title{\textit{Hubble Space Telescope} proper motion (HSTPROMO) catalogs of Galactic globular clusters$^{\ast}$. IV. Kinematic profiles and average masses of blue straggler stars}

\altaffiltext{$\ast$}{Based on proprietary and archival observations with the NASA/ESA \textit{Hubble Space Telescope}, obtained at the Space Telescope Science Institute, which is operated by AURA, Inc., under NASA contract NAS 5-26555.}

\author{A.~T.~Baldwin$^{1,2}$, L.~L.~Watkins$^2$, R.~P.~van~der~Marel$^2$, P.~Bianchini$^3$, A.~Bellini$^2$, J.~Anderson$^2$}

\affil{$^1$Department of Physics \& Astronomy, Louisiana State University, Baton Rouge, LA 70803, USA}
\affil{$^2$Space Telescope Science Institute, 3700 San Martin Drive, Baltimore MD 21218, USA}
\affil{$^3$Max Planck Institute for Astronomy, K\"{o}nigstuhl 17, D-69117 Heidelberg, Germany}
\email{abaldw8@tigers.lsu.edu;lwatkins@stsci.edu}

\begin{abstract}
    We make use of the \textit{Hubble Space Telescope} proper-motion catalogs derived by \citet{bellini2014} to produce the first radial velocity-dispersion profiles $\sigma(R)$ for blue straggler stars (BSSs) in Galactic globular clusters (GCs), as well as the first dynamical estimates for the average mass of the entire BSS population. We show that BSSs typically have lower velocity dispersions than stars with mass equal to the main-sequence turnoff mass, as one would expect for a more massive population of stars. Since GCs are expected to experience some degree of energy equipartition, we use the relation $\sigma \propto M^{-\eta}$, where $\eta$ is related to the degree of energy equipartition, along with our velocity-dispersion profiles to estimate BSS masses. We estimate $\eta$ as a function of cluster relaxation from recent Monte Carlo cluster simulations by \citet{bianchini2016b} and then derive an average mass ratio $\Mbss/\Mto = 1.50 \pm 0.14$ and an average mass $\Mbss = 1.22 \pm 0.12$~\Msun from 598 BSSs across 19 GCs. The final error bars include any systematic errors that are random between different clusters, but not any potential biases inherent to our methodology. Our results are in good agreement with the average mass of $\Mbss = 1.22 \pm 0.06$~\Msun for the 35 BSSs in Galactic GCs in the literature with properties that have allowed individual mass determination.
\end{abstract}


\section{Introduction}

Blue-straggler stars (BSSs) are hydrogen-burning stars that occupy a region of the optical color-magnitude diagram (CMD) brighter and bluer than the main-sequence turnoff (MSTO). They were first discovered in M3 by \citet{sandage1953} and have since been detected in all Galactic globular clusters (GCs). We refer to \citet{cannon2015} for a more detailed review of early BSS research.  BSSs appear to extend the main sequence into higher-mass stars, which should have evolved into giants or stellar remnants if they had formed at the same time as the rest of the cluster. BSSs mimic a younger population of stars, but GCs do not contain sufficient gas to support recent or ongoing star formation. In order to explain BSSs, then, there must be some mechanism through which pre-existing main-sequence stars can increase in mass and luminosity. This is primarily expected to occur through mass transfer in evolved binary systems \citep{sollima2008, knigge2009, geller2011, leigh2013, gosnell2014} or through stellar collisions in the cluster core \citep{hurley2005, geller2013, chatterjee2013}. Neither model can adequately produce the entire BSS population so the most likely explanation is some combination of the two \citep{ferraro2009, dalessandro2013}.

The masses of individual BSSs are not, in general, well known and their status as higher-mass stars was initially inferred solely from their position on the CMD. This hypothesis was finally put to the test by \citet{shara1997} who measured the surface gravity of an individual BSS in 47\,Tuc to derive a mass of $1.7 \pm 0.4$~\Msun; nearly twice the MSTO mass in the cluster. Individual BSS masses have since been measured for 35 stars via spectroscopic analysis \citep{demarco2005} and stellar pulsations \citep{gilliland1998, fiorentino2014}. These studies seem to suggest that a typical BSS is significantly more massive than stars at the MSTO, but not more than twice as massive, as one would expect for a population of stars formed via mass transfer or mergers of main-sequence stars. Stars significantly more massive than the MSTO will have long since evolved off of the main sequence, along with any BSSs they produced in the past. There are, of course, some exceptions to this trend such as S\,1082 \citep{vandenberg2001, sandquist2003} or WOCS\,7782 \citep{geller2009}, BSS systems that contain significantly more than twice the MSTO mass and that are likely the result of multiple stellar mergers or multiple dynamical interactions respectively, but these systems are rare and do not represent a typical BSS.

Galactic GCs have all been shown to contain BSSs and they provide a unique environment in which we can study not only BSSs, but also their dynamical interactions with the rest of the cluster. Over time, dynamical friction is expected to cause more massive objects such as BSSs to migrate towards the cluster core. This has been observed as a bimodal distribution in BSS density \citep[e.g.][]{ferraro1997} consisting of a strong central concentration, followed by a dip at intermediate radii and a subsequent rise at large radii. This distribution arises because relaxation time is a function of radius: there is some critical radius within which BSSs will have had time to migrate to the cluster core, and beyond which BSSs will have remained largely undisturbed. Another crucial result of dynamical interactions is energy equipartition. Frequent two-body interactions will tend to equalize the energy of all stars within a cluster and so more massive populations typically have lower velocity dispersions. As such, we anticipate that BSSs should have a lower velocity dispersion profile than typical main-sequence stars at a given distance from the centre. This was observed to be true for 47\,Tuc, where the BSS velocity dispersion is related to the dispersion of stars at the turnoff mass by $\sigbss / \sigto \approx \frac{1}{\sqrt{2}}$ \citep{mclaughlin2006}. This is consistent with a population of stars with twice the turnoff mass in a state of energy equipartition with the rest of the cluster.

Recently, we presented a set of proper-motion catalogues for 22 Galactic GCs, compiled from archival \textit{Hubble Space Telescope} (\textit{HST}) observations taken at multiple epochs \citep[][hereafter \citetalias{bellini2014}]{bellini2014}. Our proper-motion catalogs have several advantages over radial-velocity surveys: since proper motions are measured by determining how far stars have moved from one epoch to another, we are able to measure a large sample of stars from imaging for which it would be prohibitively time consuming to acquire the individual spectra needed for a radial-velocity survey. Further, we are able to observe fainter stars for which reliable spectra may not be available at all. In \citet[][hereafter \citetalias{watkins2015a}]{watkins2015a}, we used the catalogs to study the radial velocity-dispersion and velocity-anisotropy profiles of all the bright stars in 22 Galactic GCs. In \citet{watkins2015b}, we compared the bright-star dispersion profiles against literature line-of-sight dispersion profiles to estimate dynamical distances and mass-to-light ratios. These studies are all part of the \textit{Hubble Space Telescope} Proper Motion (HSTPROMO) collaboration \citep{vandermarel2014}.

Here, we present the first large-scale kinematic survey of BSSs in Galactic GCs using the \citetalias{bellini2014} proper-motion catalogs. The paper is laid out as follows: \autoref{sect:data} introduces a series of cuts to our catalog, first to produce a sample of stars with a narrow range of masses and second to select the BSS population. In \autoref{sect:results}, we divide our BSS population into radial bins, use a maximum-likelihood method to estimate the velocity dispersion of each bin, and fit a dispersion profile to the binned velocity dispersion estimates. In \autoref{sect:discussion}, we estimate the typical mass of BSSs in each cluster and of BSSs as a whole, and make comparisons with previous results. Our results are summarized in \autoref{sect:conclusions}.

\section{Cluster Data}
\label{sect:data}

Although \citetalias{bellini2014} presented \textit{HST} proper-motion catalogs for 22 Galactic GCs, in this paper we make use of the catalogs for only 19 of these clusters. NGC\,6535 and NGC\,7099 (M30) are excluded from our analysis as our catalogs contain relatively few stars for these clusters and we do not detect enough BSSs in either cluster to make a meaningful estimate of their velocity dispersion. We also exclude NGC\,6715 (M54) due to the risk of contamination from the Sagittarius dwarf spheroidal galaxy. Some characteristic properties are provided in \autoref{sect:results} (\autoref{table:results}) for the 19 clusters used for this study.

These catalogs measure \textit{relative} proper motions rather than absolute proper motions. This is due to a lack of ``fixed'' background sources bright enough to be observed through the dense core region of a GC. Consequently, the average velocity of cluster members in any small region of the sky should be zero by design and these catalogs cannot be used to measure the differential rotation or bulk motion of the cluster as a whole. In this paper, we are only concerned with measuring velocity dispersions, which can be calculated from relative proper motions. We refer back to \citetalias{watkins2015a} for a more detailed explanation of these issues.

We select our BSSs from the high-quality bright-star samples described in \citetalias{watkins2015a}. Here we briefly describe the bright-star samples before explaining our BSS selection procedure.

\subsection{Bright-Star Catalog}
\label{sect:brightcat}

Accurate kinematic estimates require high-quality velocity measurements with reliable uncertainties. Including stars for which the positions are poorly determined (often due to blending with a neighbouring star) or for which the velocity uncertainties have been underestimated tends to artificially increase the velocity distributions. Contaminants -- i.e. stars that are not members of the cluster -- can also introduce biases.

To avoid such sources of error, we do not use the full catalogs from \citetalias{bellini2014}, but instead the cleaned samples of bright stars from \citetalias{watkins2015a}. We refer to Section~2 of \citetalias{watkins2015a} for further details regarding these cuts, but briefly summarize them here: To select these samples, we started with a magnitude cut at 1 magnitude below the MSTO to select only bright stars.  Next, a series of quality cuts were made on: 1) the number of individual measurements used for the proper-motion estimate; 2) the quality of the proper-motion fits; and 3) the quality of the point-spread-function fits. Finally, velocity outliers and stars with large velocity uncertainties were removed.

In \citetalias{watkins2015a}, we were interested only in radial changes of the kinematics and wished to neglect the effects of stellar mass. The magnitude cut was made to restrict the range of stellar mass in each cluster sample. As we will see, this cut is fainter than the faint magnitude limit we will impose on our BSS samples, so it will not interfere with our selection.

\subsection{Blue-Straggler Selection}

BSSs are an apparent extension of the main sequence, both brighter and bluer than the MSTO, so we must make cuts in both color and magnitude to separate them from the rest of the bright-star catalog. We first identify the MSTO as follows: we bin all of the stars in our catalog into bins 0.1~mag wide, fit a Gaussian to the color distribution of each bin, and take the bin with the bluest mean to be the MSTO \citepalias[see also Section~2.1 of][]{watkins2015a}. We then identify the color and color dispersion of the MSTO by selecting all of the stars within 0.1~mag of the MSTO and calculating both the 5$\sigma$-clipped mean color and its standard deviation, which we denote as $\sigc$.

\begin{table}
    \caption{Color and magnitude cuts for each cluster.}
    \label{table:bsscuts}
    \centering
    
    \begin{tabular}{cccccccc}
        \hline
        \hline
        Cluster & Faint cut  & Bright cut & Red cut & Blue cut & $N_\mathrm{BSS}$ \\
        & (mag) & (mag) & (mag) & (mag) & \\
        (1) & (2) & (3) & (4) & (5) & (6) \\
        \hline
        NGC\,104  & 16.9 & 14.8 & 0.56 & 0.28 & 25 \\
        NGC\,288  & 18.4 & 16.4 & 0.52 & 0.10 & 27 \\
        NGC\,362  & 18.2 & 16.5 & 0.50 & 0.10 & 40 \\
        NGC\,1851 & 18.9 & 17.2 & 0.51 & 0.10 & 24 \\
        NGC\,2808 & 18.8 & 16.6 & 0.67 & 0.20 & 58 \\
        NGC\,5139 & 17.9 & 15.7 & 1.090 & 0.20 & 73 \\
        NGC\,5904 & 17.8 & 17.0 & 0.51 & 0.10 & 16 \\
        NGC\,5927 & 18.8 & 16.5 & 0.97 & 0.60 & 65 \\
        NGC\,6266 & 18.8 & 16.5 & 2.15 & 1.00 & 19 \\
        NGC\,6341 & 18.0 & 16.0 & 0.46 & 0.00 & 37 \\
        NGC\,6362 & 18.0 & 16.0 & 0.58 & 0.27 & 21 \\
        NGC\,6388 & 19.6 & 17.5 & 0.91 & 0.40 & 58 \\
        NGC\,6397 & 15.7 & 14.5 & 0.64 & 0.07 & 10 \\
        NGC\,6441 & 20.2 & 18.0 & 1.03 & 0.60 & 25 \\
        NGC\,6624 & 18.7 & 16.8 & 0.80 & 0.50 & 12 \\
        NGC\,6656 & 16.8 & 15.6 & 0.85 & 0.46 & 34 \\
        NGC\,6681 & 18.5 & 17.0 & 0.58 & 0.10 & 14 \\
        NGC\,6752 & 16.6 & 15.0 & 0.52 & 0.00 & 16 \\
        NGC\,7078 & 18.6 & 17.3 & 0.52 & 0.10 & 21 \\
        \hline
    \end{tabular}
    
    \qquad
    
    \raggedright
    \textbf{Notes.} Columns: (1) cluster identification (2) faint magnitude cut; (3) manual bright magnitude cut; (4) red color cut; (5) manual blue color cut; (6) number of BSSs used in our analysis. In some cases, the number of BSSs used is lower than the number detected in our catalog since some BSS were too isolated to be sensibly binned for dispersion estimates. For most clusters, the magnitudes given are F814W and colors are defined as F606W-F814W. For NGC\,5139, the given magnitude is F625W and the color is defined is F435W-F625W. For NGC\,6266, the given magnitude is F658N and the color is defined as F390W-F658N.
\end{table}

Now that we have characterized the MSTO, we are ready to select BSSs. We first select stars that are at least 0.1~mag brighter than the MSTO. This number is chosen as we have only constrained the MSTO to within a 0.1~mag bin, which is large relative to the photometric uncertainty of our measurements. This cut is sufficient to ensure that only stars brighter than the MSTO appear in our BSS catalog.

Next, we select for stars that are bluer than the MSTO. A binary or multiple-star system of main-sequence stars near the MSTO could mimic a BSS. To account for this, we select only stars that are at least 3$\sigc$ bluer than the MSTO. A binary or multiple-star system will appear brighter on the CMD, but it will not appear any bluer so this cut should be sufficient to ensure that we select only stars that are truly distinct from the main sequence.

After making these cuts, we often need to make additional cuts in both color and magnitude to remove the horizontal branch from our BSS sample. These extra cuts are made by eye on a cluster-by-cluster basis. All cuts are given in \autoref{table:bsscuts}.

\begin{figure}
    \centering
    \includegraphics[width=\linewidth]{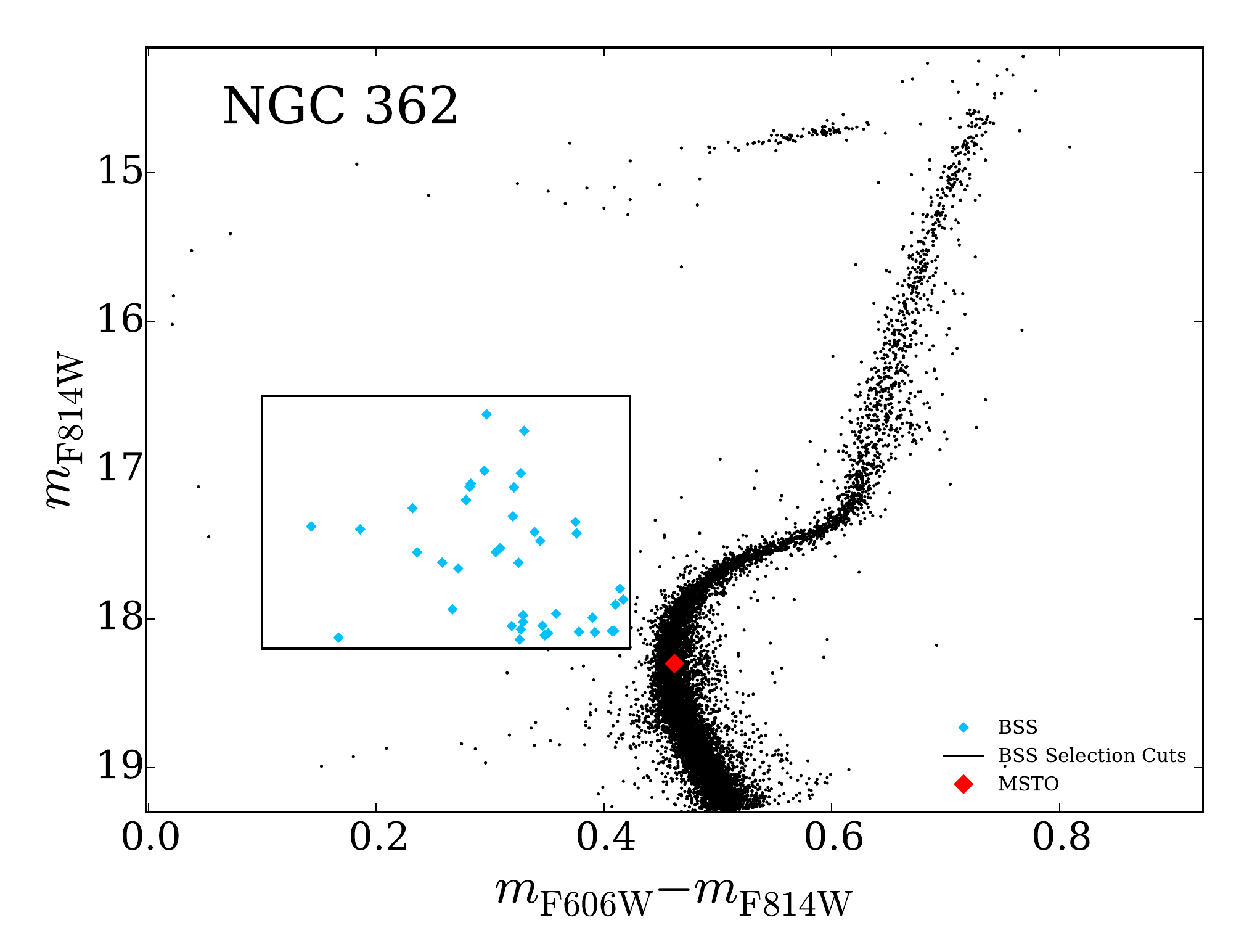}
    \caption{CMD for NGC\,362, illustrating our BSS selection. The black points show stars from the bright-star catalog, the blue diamonds show the selected BSSs, and the red diamond marks the adopted position of the MSTO. The black lines show the cuts made to isolate the BSS population.}
    \label{fig:cmd_example}
\end{figure}

\begin{figure*}
    \centering
    \includegraphics[width=0.33\linewidth]{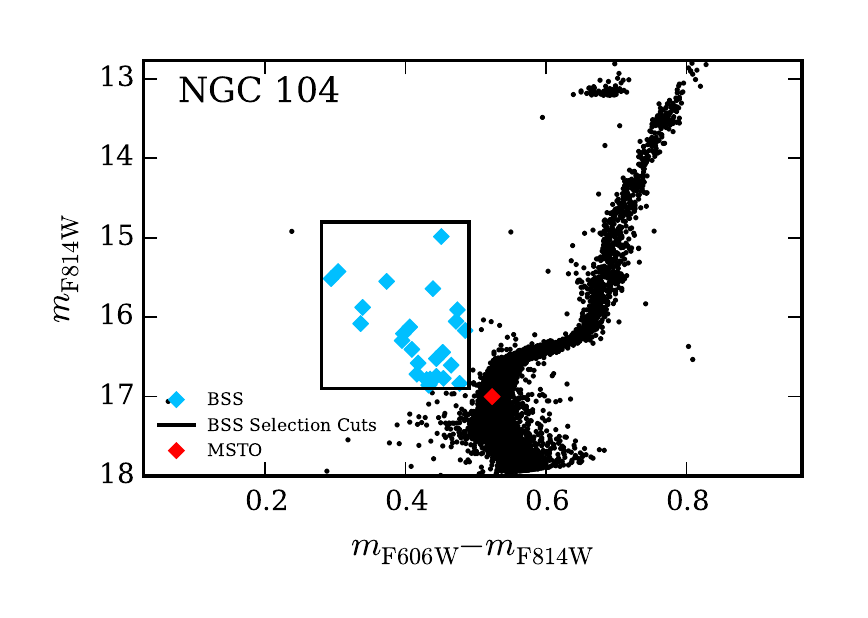}
    \includegraphics[width=0.33\linewidth]{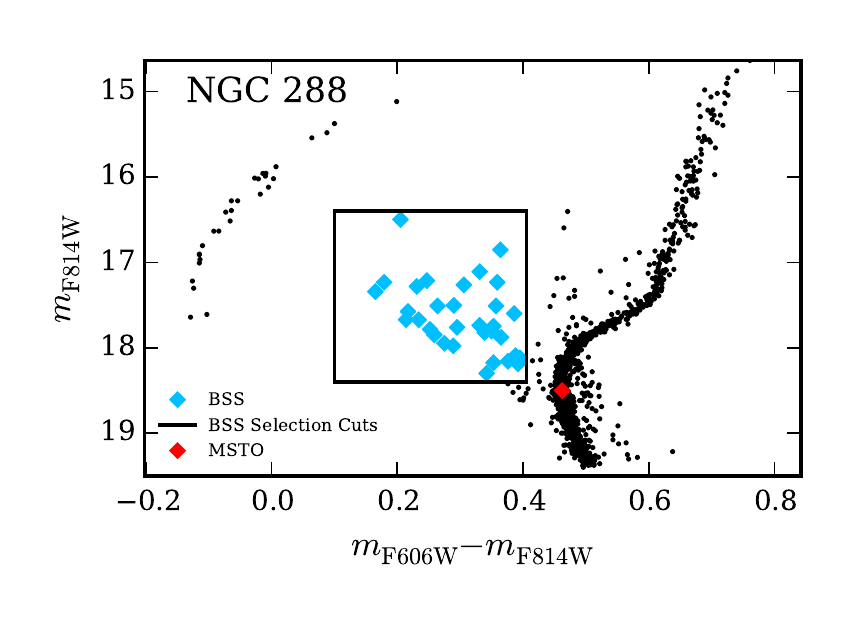}
    \includegraphics[width=0.33\linewidth]{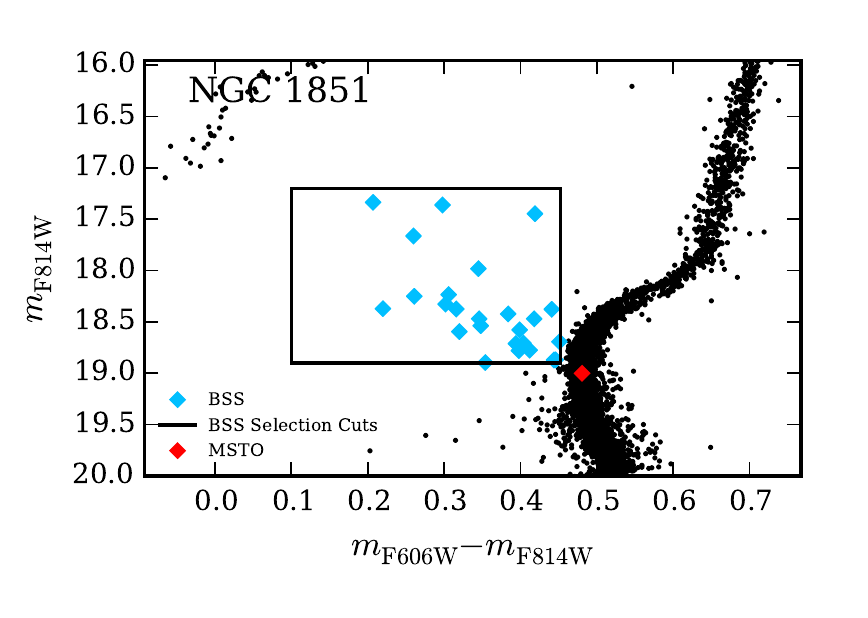}
    \includegraphics[width=0.33\linewidth]{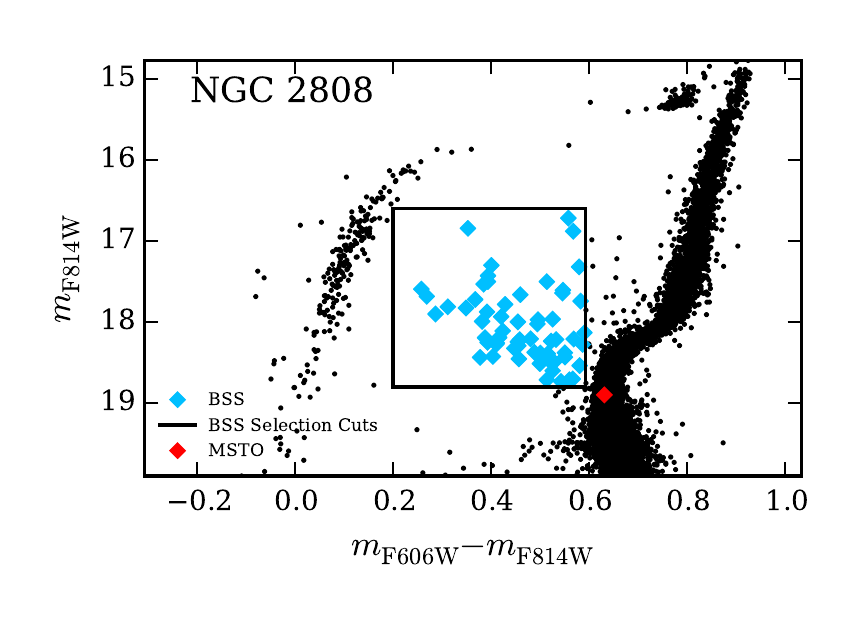}
    \includegraphics[width=0.33\linewidth]{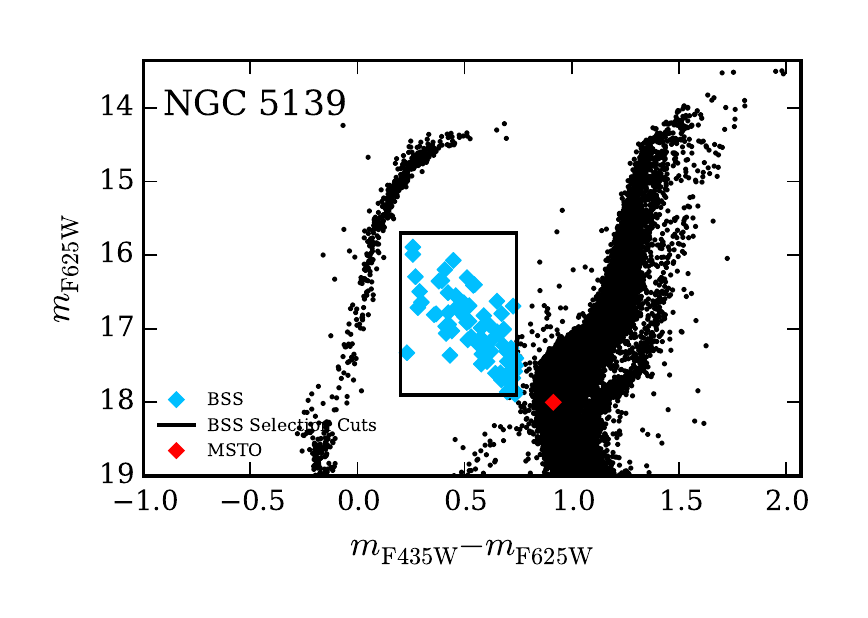}
    \includegraphics[width=0.33\linewidth]{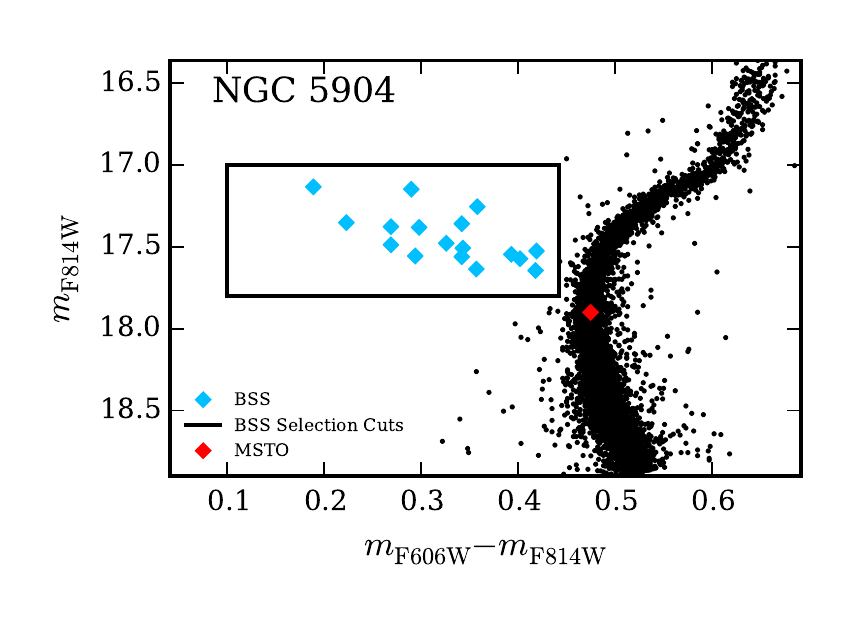}
    \includegraphics[width=0.33\linewidth]{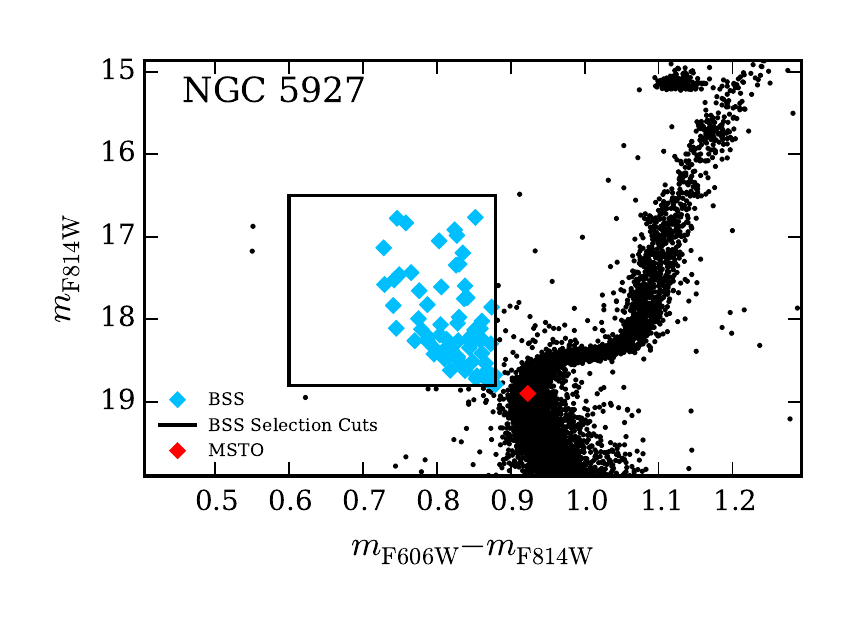}
    \includegraphics[width=0.33\linewidth]{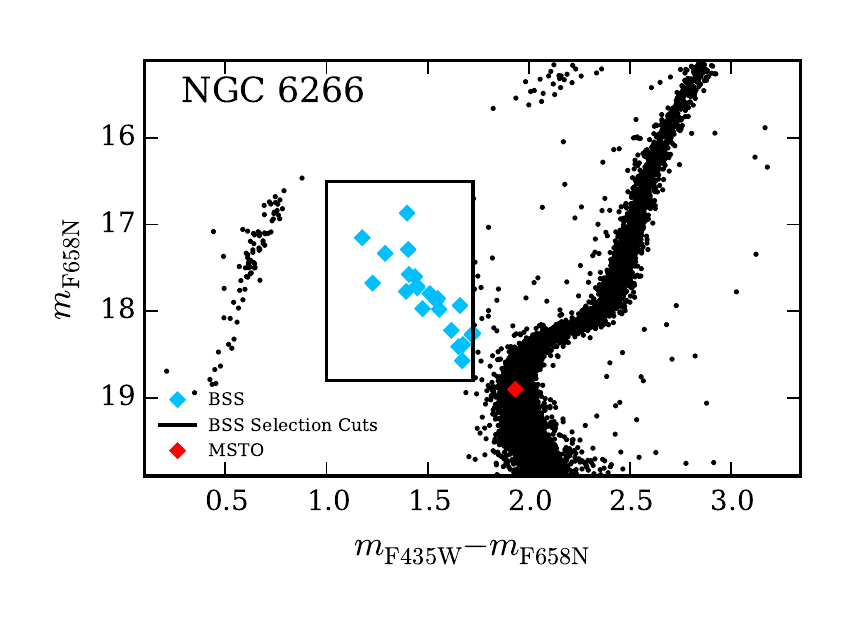}
    \includegraphics[width=0.33\linewidth]{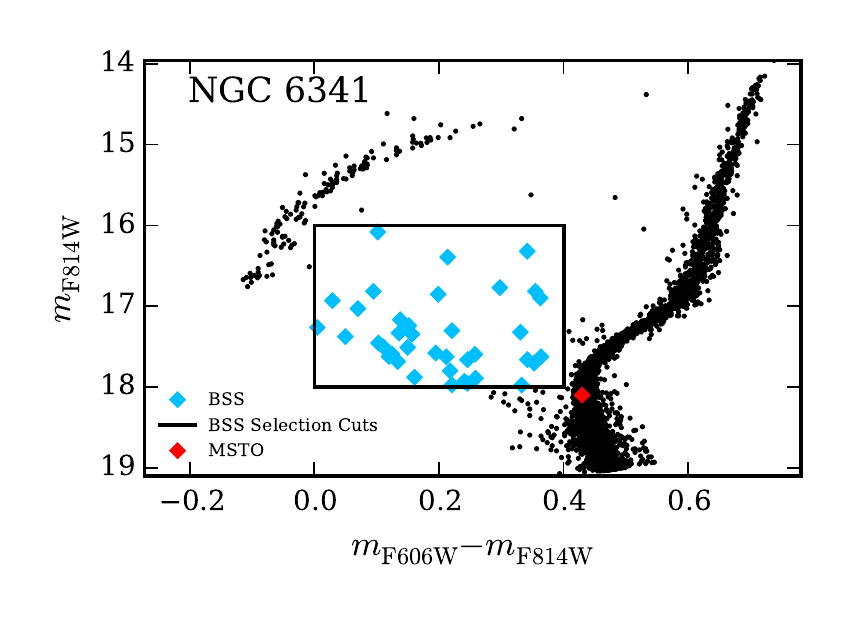}
    \caption{CMDs for NGC\,104, NGC\,288, NGC\,1851, NGC\,2808, NGC\,5139, NGC\,5904, NGC\,5927, NGC\,6266 and NGC\,6341 with the relevant color and magnitude cuts to isolate BSSs. See the caption to \autoref{fig:cmd_example} for more details.}
    \label{fig:cmds1}
\end{figure*}

\begin{figure*}
    \centering
    \includegraphics[width=0.33\linewidth]{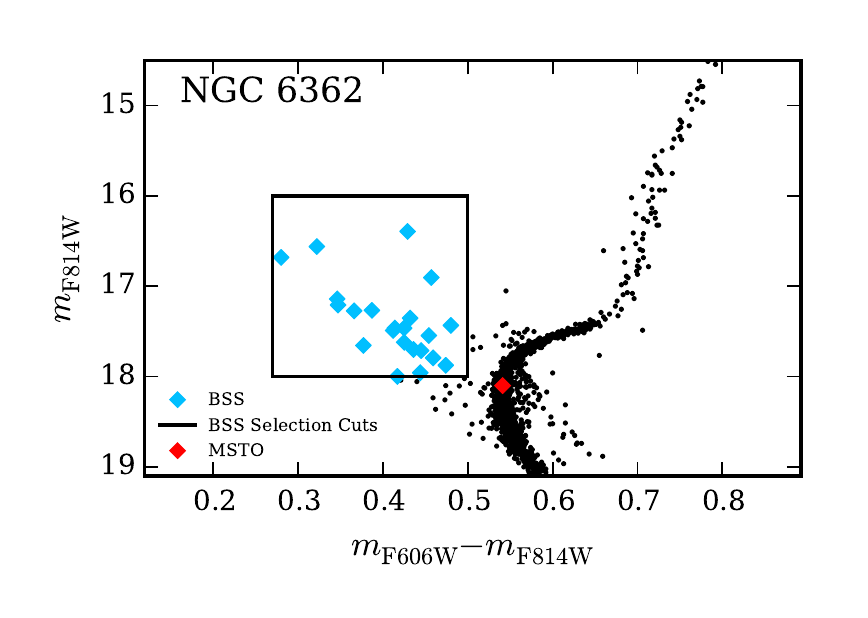}
    \includegraphics[width=0.33\linewidth]{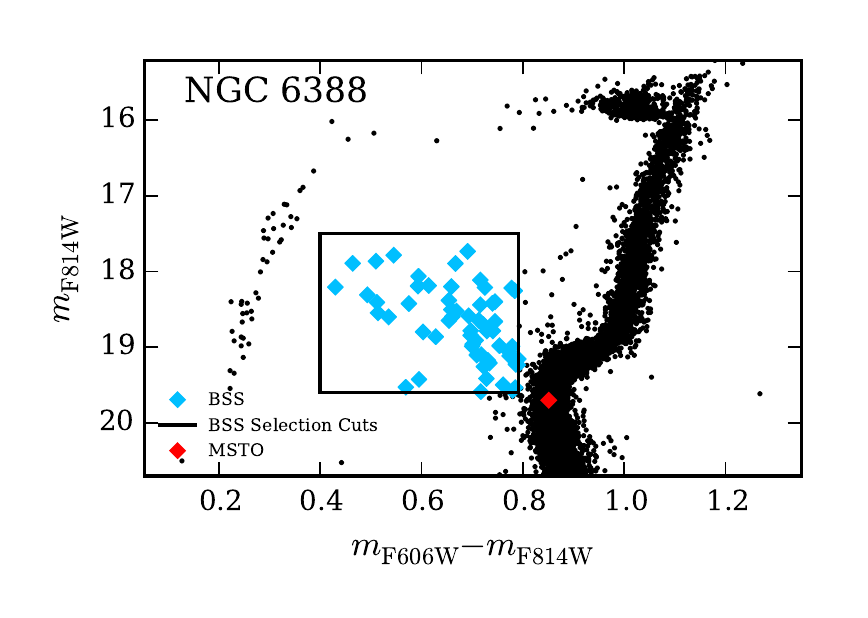}
    \includegraphics[width=0.33\linewidth]{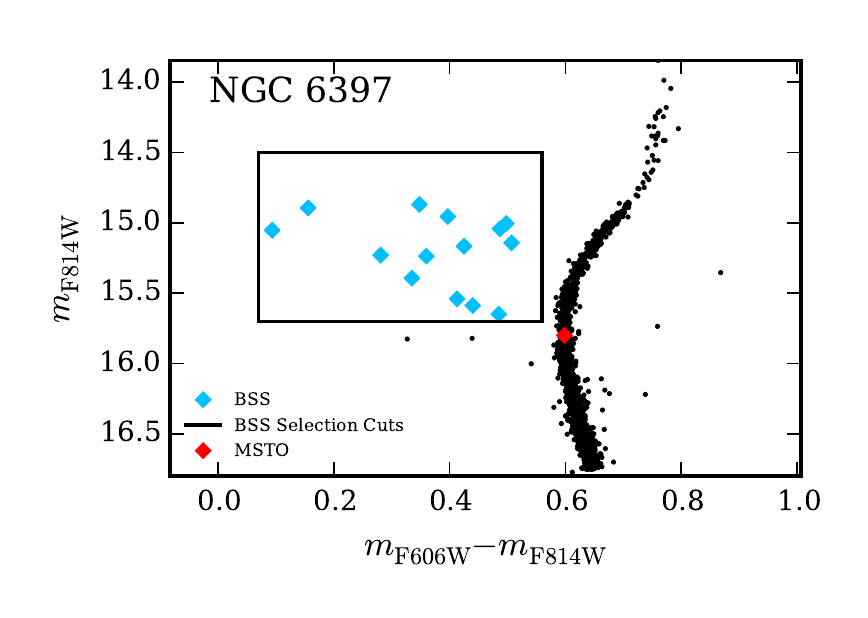}
    \includegraphics[width=0.33\linewidth]{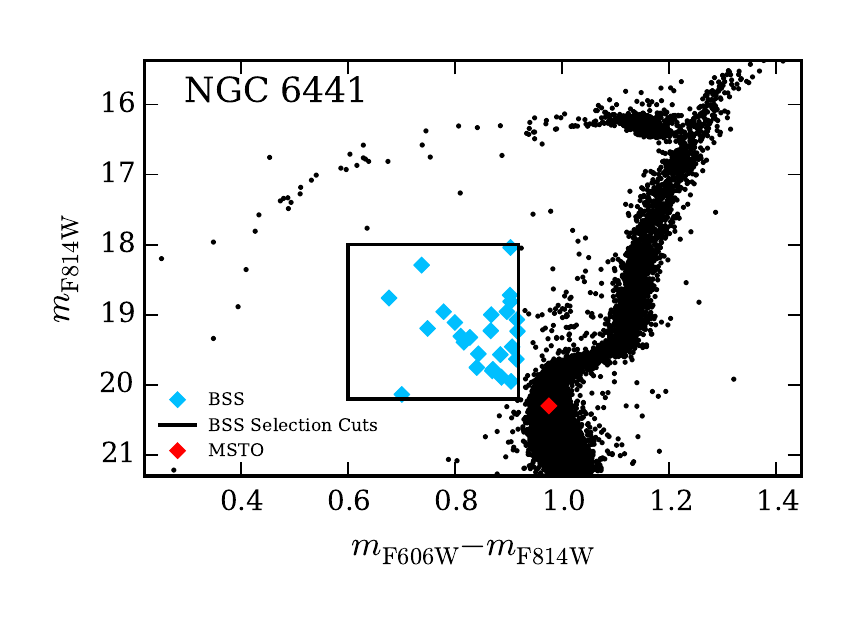}
    \includegraphics[width=0.33\linewidth]{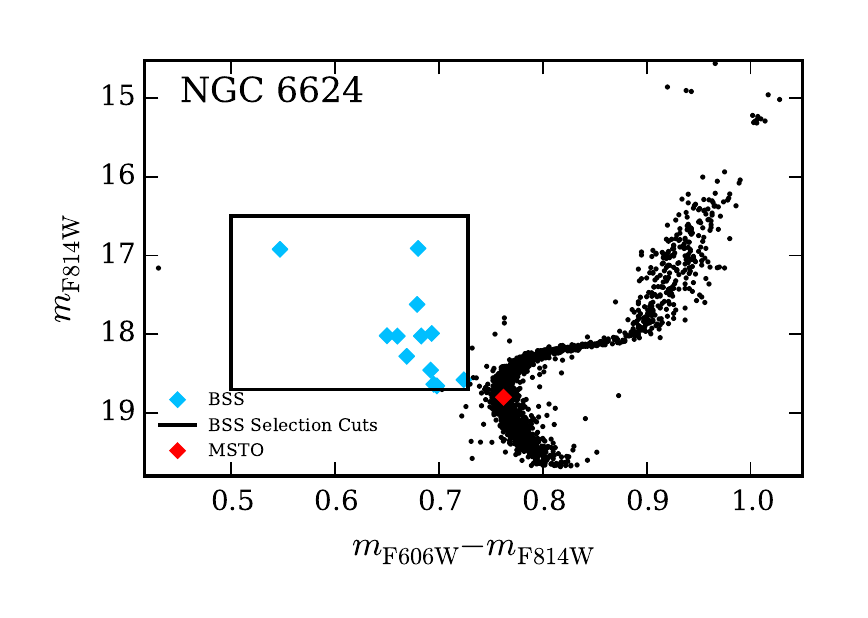}
    \includegraphics[width=0.33\linewidth]{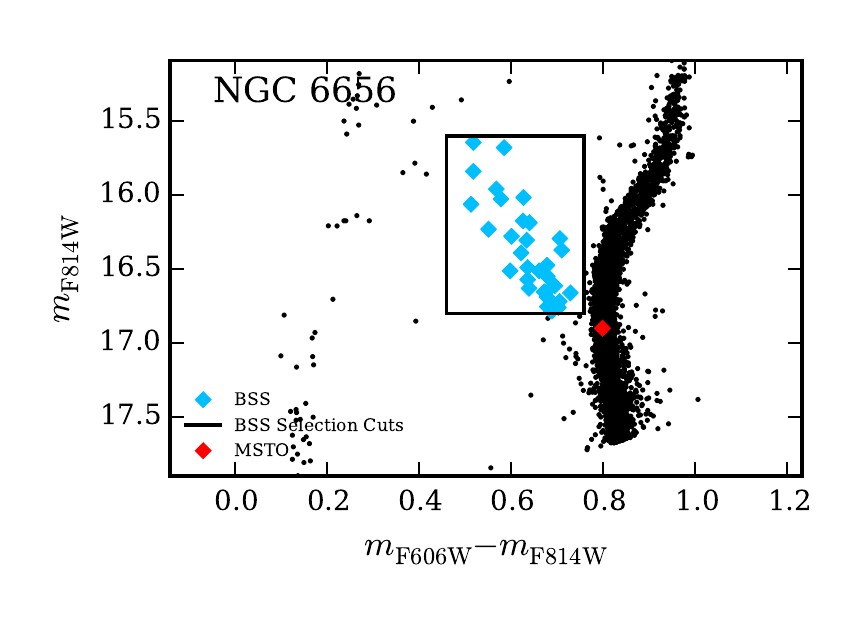}
    \includegraphics[width=0.33\linewidth]{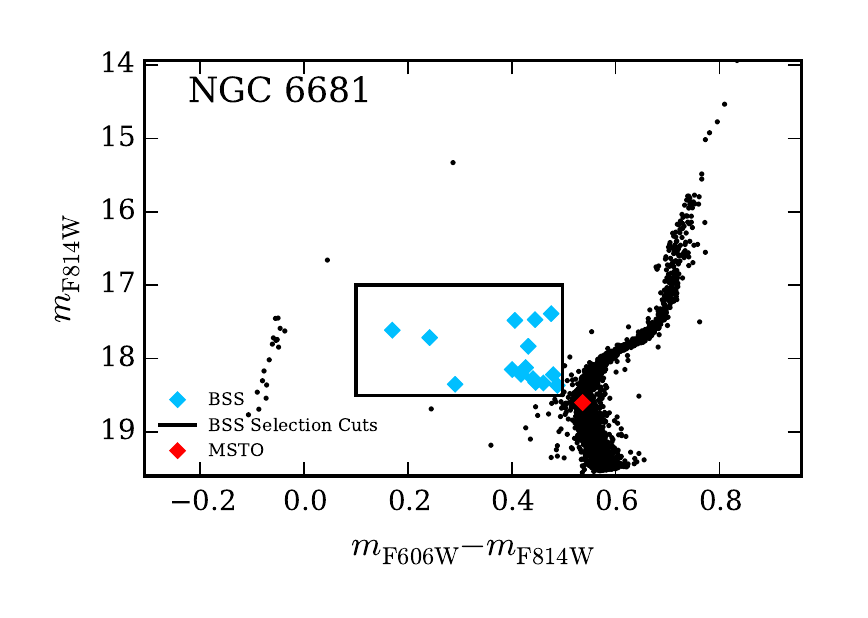}
    \includegraphics[width=0.33\linewidth]{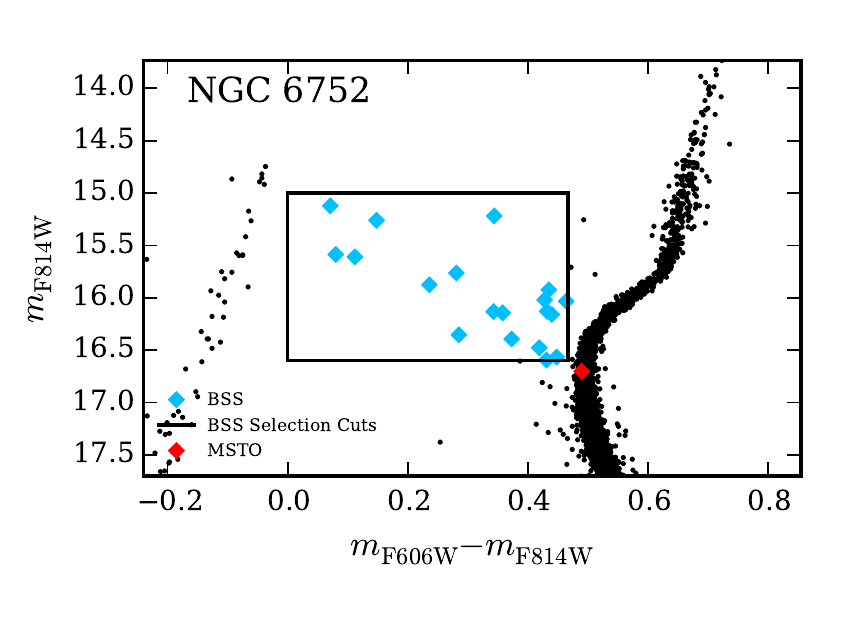}
    \includegraphics[width=0.33\linewidth]{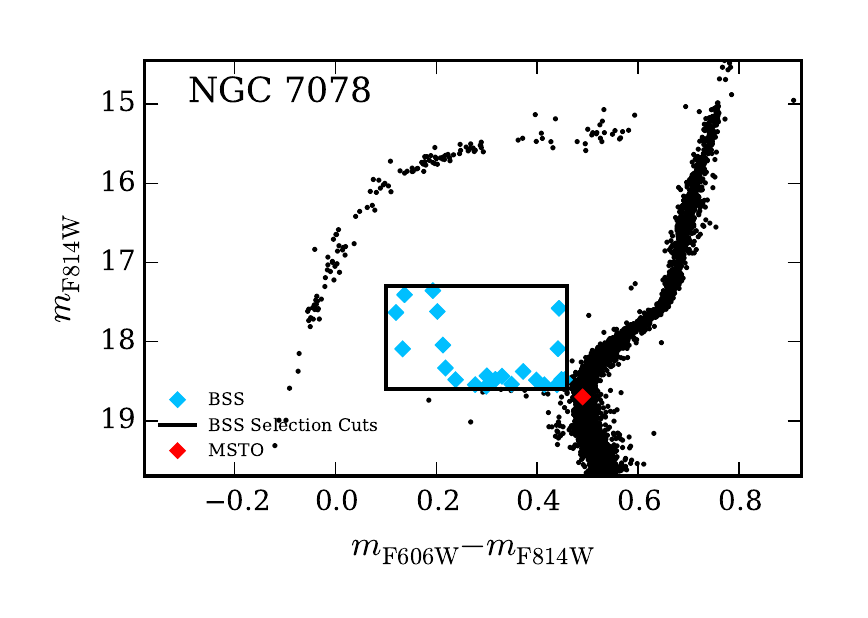}
    \caption{CMDs for NGC\,6362, NGC\,6388, NGC\,6397, NGC\,6441, NGC\,6624, NGC\,6656, NGC\,6681, NGC\,6752 and NGC\,7078 with the relevant color and magnitude cuts to isolate BSSs. See the caption to \autoref{fig:cmd_example} for more details.}
    \label{fig:cmds2}
\end{figure*}

As an example, we show the CMD for NGC\,362 in \autoref{fig:cmd_example} with our selected BSSs shown as blue diamonds and all other stars plotted as black points. The red diamond marks the adopted location of the MSTO. The black lines mark our BSS selection cuts. Figures~\ref{fig:cmds1} and \ref{fig:cmds2} present the CMDs for the rest of the clusters in our sample.

\section{Results}
\label{sect:results}

Our aim is to compare the radial velocity-dispersion profiles of BSSs with the radial velocity-dispersion profiles of stars with masses on the order of the turnoff mass. In \citetalias{watkins2015a}, we calculated velocity-dispersion profiles for the bright stars in each cluster. To each we fit a monotonically-decreasing fourth-order polynomial that was defined to be flat at small radii. These polynomial fits are described in detail and displayed in \citetalias{watkins2015a}. The best fits for each cluster are used in this paper as a morphological model for our BSS dispersion profiles, as we will later discuss.

Two-body interactions between stars are known to preferentially equalize the kinetic energy of the two stars. That is to say, kinetic energy is most-often transferred from a higher-energy star to a lower-energy star. Stars within the ancient and densely-populated environment of a GC will have experienced many such interactions during their lifetime and, as a result, we expect them to evolve towards a state where all stars in the cluster have the same kinetic energy. This state is called energy equipartition. The kinetic energy of a population of $N$ stars of mass $M$, average velocity $\bar{v} = 0$, and velocity dispersion $\sigma$ is proportional to $\sum_{i=1}^{N} $ $M\sigma^2$. So, for a system in complete energy equipartition, we would expect $\sigma(M) \propto M^{-0.5}$. In practice, GCs are not found in complete energy equipartition \citep{trenti2013, anderson2010}, and instead follow the power law,
\begin{equation}
    \sigma(M) \propto M^{-\eta},
    \label{eqn:sigmeta}
\end{equation}
where $\eta$ is a constant between 0 and 0.5 that depends on the type of system in question and the parameters of the GC as a whole (e.g. core concentration, relaxation time).\footnote{As we will discuss in \autoref{sect:eta}, this power-law approximation is generally only valid for limited mass ranges but does not hold globally for GCs \citep[see also][]{bianchini2016b}.}

Recall from \autoref{sect:brightcat} that we have limited ourselves to a catalog of bright stars, where `bright' is defined as no more than 1~mag below the MSTO. This is advantageous for our present analysis because it represents a narrow range of stellar masses, which allows us to produce standard profiles against which we can compare our BSS profiles.

Previous work \citep[eg.][]{sabbi2004, ferraro2006, dalessandro2008} has shown that the relative radial distributions of BSSs in clusters can be flat, bimodal, or centrally peaked, depending on the dynamical state of the cluster.\footnote{Typically, the radial distribution of BSS stars is compared to the radial distribution of some reference population, such as horizontal-branch or red-giant-branch stars.} In the case of a bimodal radial distribution, it is possible that core stragglers could have different properties from the outer stragglers. However, as our BSS samples are primarily from the cluster cores, we do not consider such differences here.

\subsection{Blue-Straggler Dispersion Profiles}
\label{sect:bssdisps}

To estimate BSS kinematic profiles for each cluster, we bin each BSS population in radius and then estimate the velocity dispersion in each bin using the maximum-likelihood method described in Section~3.1 of \citetalias{watkins2015a}. Due to the relative rarity of BSSs, the radial distribution of BSS varies significantly between clusters and we are unable to apply a single binning algorithm to all of the clusters within our sample. Each cluster is, therefore, manually binned in an effort to minimize the radial extent of each bin while maximizing the number of stars per bin and the number of available bins. Ideally, we try to make our bins small enough that the dispersion profile should not vary appreciably across the radial extent of the bin but this is not always possible for sparsely populated clusters.

BSSs make up a tiny fraction of each cluster; our catalogs have fewer than 75 BSSs, even for the most-populated clusters in our sample. We reduce the minimum number of stars in each bin from 25 \citepalias[as used in][]{watkins2015a} to 7 to increase the spatial resolution of our dispersion profiles. The fractional error on our velocity dispersion measurement is given by
\begin{equation}
    \frac{\Delta \sigma}{\sigma} = \sqrt{\frac{1}{2N_\mathrm{v}}} ,
\end{equation}
where $N_\mathrm{v}$ is the number of velocity measurements. In this case, $N_\mathrm{V}$ is twice as large the number of stars in each bin since we have both radial and tangential proper motions for each star. With a bin size of 7, we achieve a fractional error less than 0.2, which is reasonable for our purposes.

It is difficult to constrain the shape of a BSS dispersion profile based on a small number of radial bins. Instead, we use the fact that, as a system moves towards energy equipartition, we expect velocity dispersion to vary as a function of mass, following \autoref{eqn:sigmeta}. We therefore assume for simplicity that the BSSs follow a dispersion profile morphologically similar to stars of the turnoff mass modified by some factor,
\begin{equation}
    \alpha \equiv \frac{\sigbss}{\sigto} = \left( \frac{\Mbss}{\Mto}
    \right)^{-\eta},
    \label{eqn:alpha}
\end{equation}
where $\sigbss$ and $\Mbss$ are the dispersion and mass of the BSSs, $\sigto$ and $\Mto$ are the dispersion and mass of the turnoff stars, and $\eta$ represents the degree of energy equipartition experienced by BSSs in the cluster, which for simplicity is assumed here to be independent of radius.

Let us define $f(R)$ to be the polynomial fit to the bright-star dispersion from \citetalias{watkins2015a}. Then for any given scale factor $\alpha$, the model dispersion profile is $\sigma(R) = \alpha f(R)$. For a bin $i$ at position $\Ri$, the likelihood $\Li$ of the observed velocity dispersion $\sigi \pm \dsigi$ given the model prediction is,
\begin{align}
    \Li & = p \left( \sigi | \alpha, \Ri, \dsigi \right) \nonumber \\
    & = \frac{1}{\sqrt{2 \pi \dsigi^2}} 
    \exp \left[-\frac{\left( \sigi - \alpha f(R_\mathrm{i}) \right)^2}{2 \dsigi^2} \right].
\end{align}
The posterior probability $\Posti$ of the model $\alpha$ given the observed properties of bin $i$ is then,
\begin{align}
    \Posti & = p \left( \alpha | \sigi, \Ri, \dsigi \right) \nonumber \\
    & = p \left( \sigi | \alpha, \Ri, \dsigi \right) p \left( \alpha \right) \nonumber \\
    & = \Li p \left( \alpha \right),
\end{align}
where $p \left( \alpha \right)$ is the prior probability of $\alpha$, which we will assume is constant. We wish to find the value of $\alpha$ that maximises the total posterior for all $N$ bins, however, as we are assuming a flat prior on $\alpha$, we need only maximise the total likelihood $\mathcal{L}$,
\begin{equation}
    \mathcal{L} = \prod_{i=1}^{N} \Li.
\end{equation}

We use \textsc{emcee}, an affine-invariant Markov Chain Monte Carlo (MCMC) sampler \citep{foreman-mackey2013}, to explore the parameter space and sample the region of best fit. We use 250 trial points (walkers) per step, and find 200 steps to be sufficient for our walkers to converge. We take the final position of each of our 250 walkers to represent a family of fits to our data. This method returns an approximately Gaussian distribution of values for $\alpha$; we take the mean to be our estimate for $\alpha$ and the dispersion to be our $1\sigma$ error estimate.

\begin{figure}
    \centering
    \includegraphics[width=0.9\linewidth]{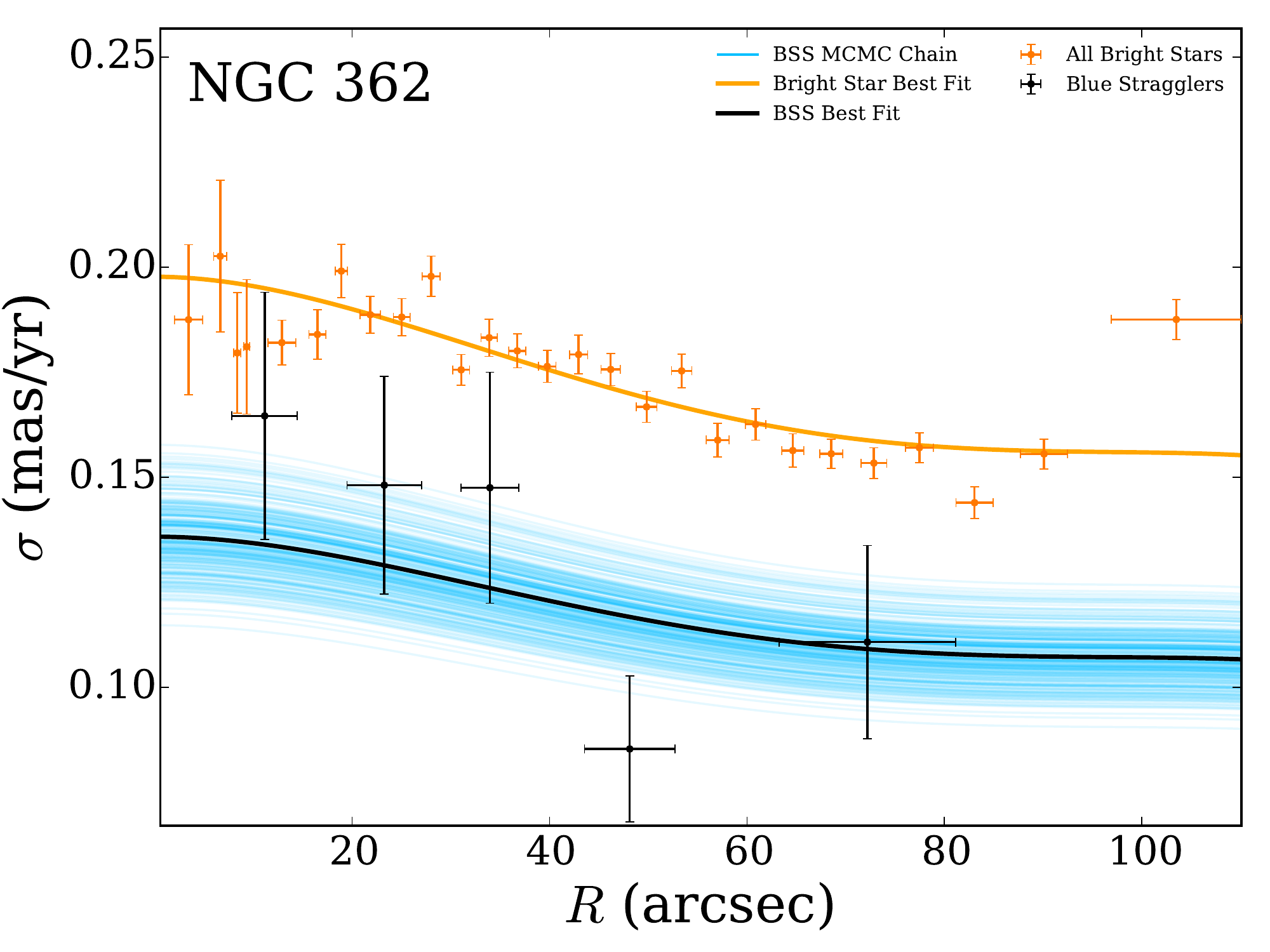}
    \caption{Velocity dispersion profile for NGC\,362. The orange points show the binned dispersion profile for all bright stars (with masses of the order of the MSTO mass) from \citetalias{watkins2015a}. The black points show the binned dispersion estimates for the BSS sample; these points clearly fall below the bright-star profile, as expected for a population of higher mass stars in a system with some degree of energy equipartition. The orange line shows a polynomial fit to the bright-stars \citepalias[also from][]{watkins2015a}. To the BSS points, we fit a profile with the same shape as the bright-star profile, but scaled by a factor $\alpha = \left( \Mbss / \Mto \right)^{-\eta}$ to estimate the relative mass difference. The blue lines show draws from the MCMC fit; the adopted `best' fit is shown in black.}
    \label{fig:disp_example}
\end{figure}

We show the radial velocity-dispersion profile for NGC\,362 in \autoref{fig:disp_example}. The orange points represent the binned dispersion estimates for all bright stars from \citetalias{watkins2015a} and the black points show our binned dispersion estimates for BSSs. In this case, the BSS dispersion profile clearly falls well below the dispersion profile for all bright stars, as we would expect for a more massive population of stars in a system approaching energy equipartition. The orange line shows the best-fit polynomial to the bright stars \citepalias[also from][]{watkins2015a}. We use this to represent the dispersion profile for stars with mass comparable to the turnoff mass and then scale the profile to fit the BSS profile. The blue lines show a family of fits to the data obtained from the MCMC sampling. The adopted `best' estimate is shown in black.

\begin{figure*}
    \centering
    \includegraphics[width=0.33\linewidth]{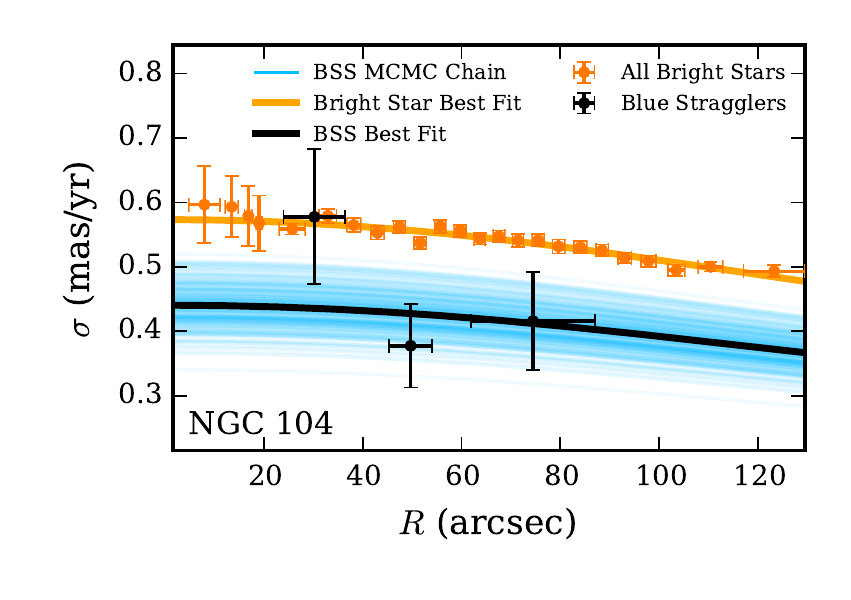}
    \includegraphics[width=0.33\linewidth]{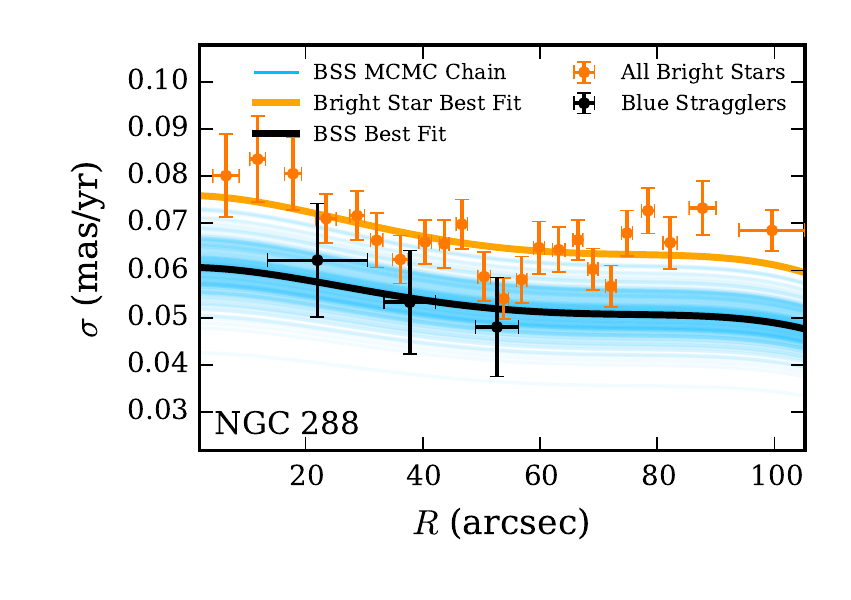}
    \includegraphics[width=0.33\linewidth]{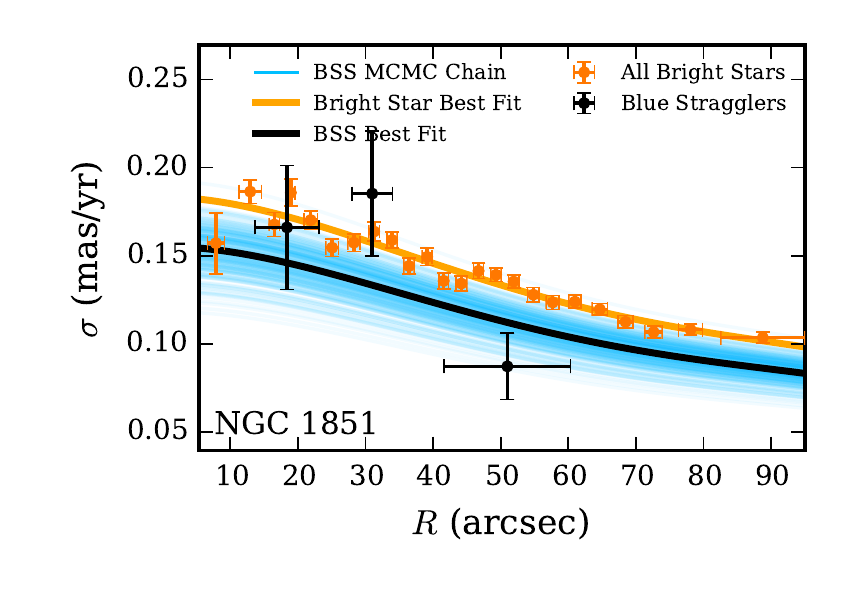}
    \includegraphics[width=0.33\linewidth]{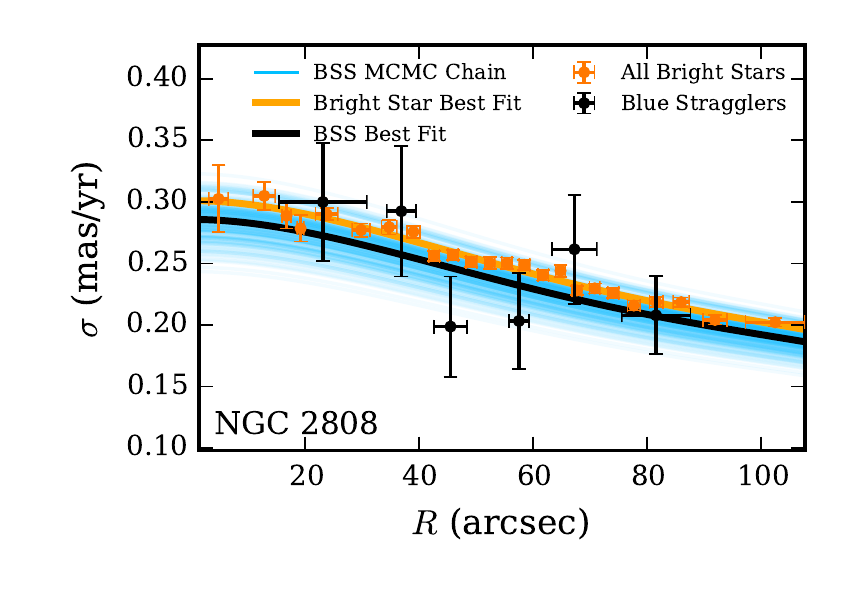}
    \includegraphics[width=0.33\linewidth]{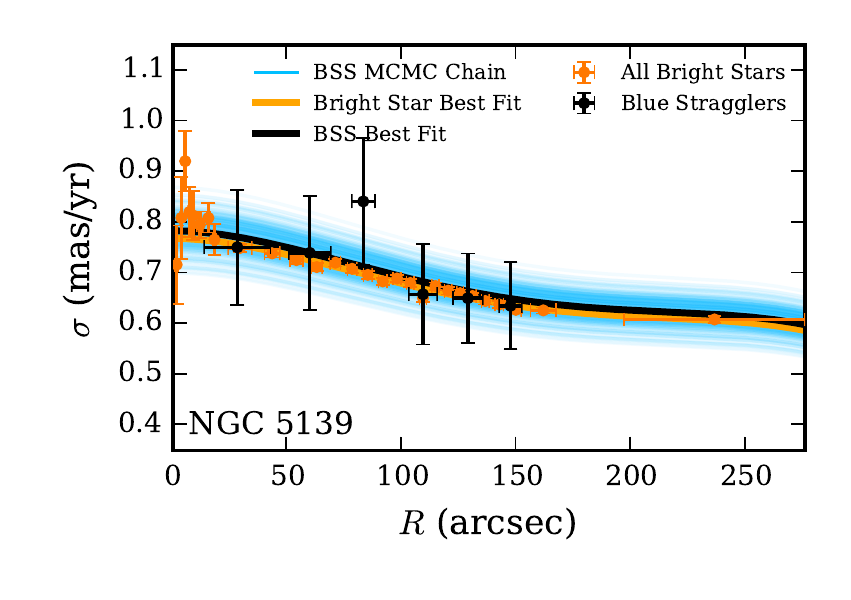}
    \includegraphics[width=0.33\linewidth]{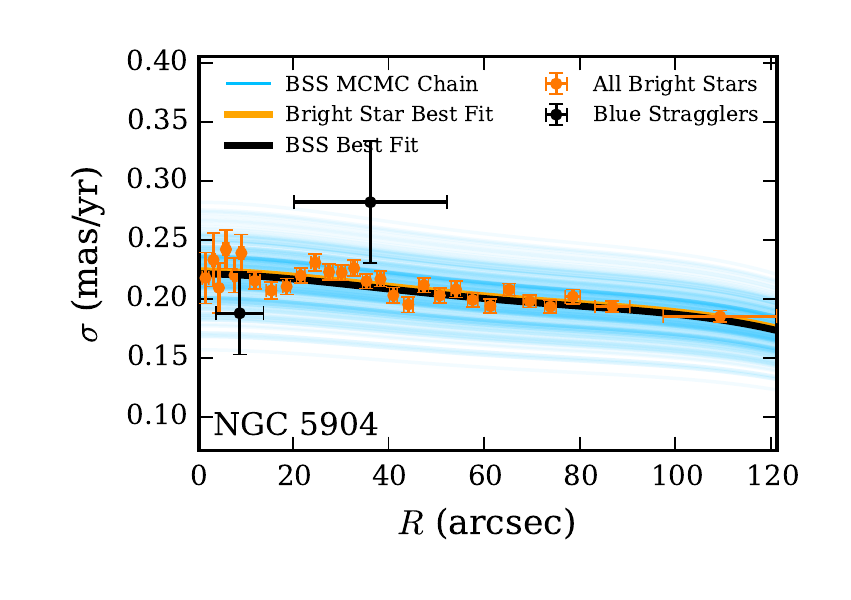}
    \includegraphics[width=0.33\linewidth]{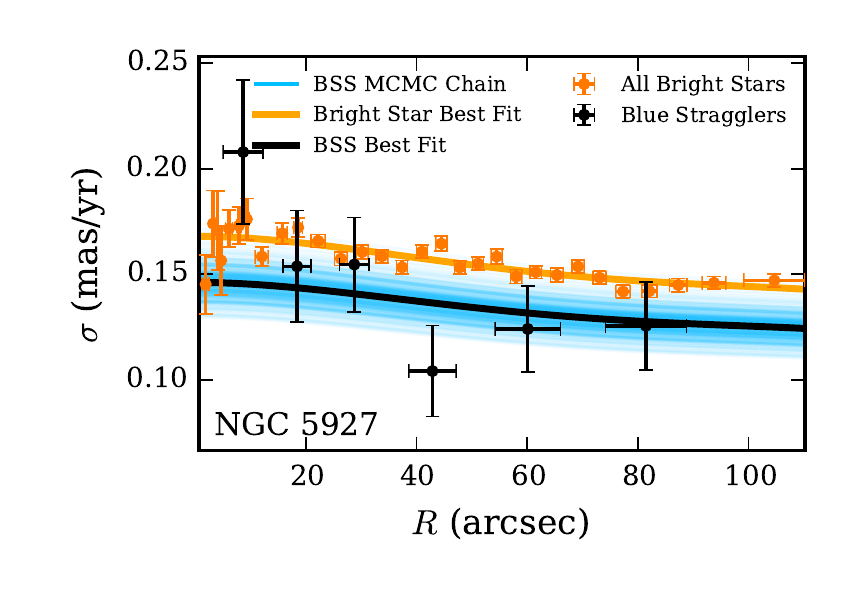}
    \includegraphics[width=0.33\linewidth]{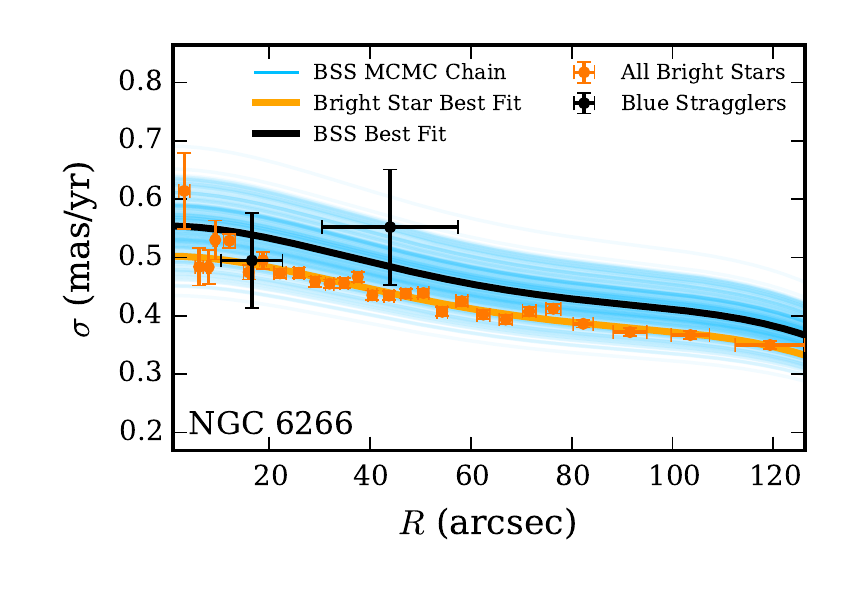}
    \includegraphics[width=0.33\linewidth]{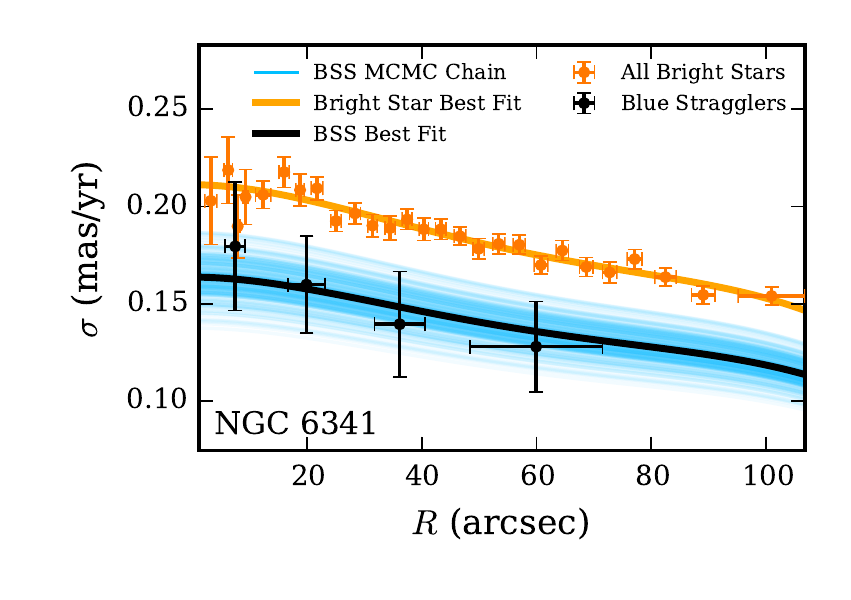}
    \caption{Velocity dispersion profiles for NGC\,104, NGC\,288, NGC\,1851, NGC\,2808, NGC\,5139, NGC\,5904, NGC\,5927, NGC\,6266 and NGC\,6341. See the caption to \autoref{fig:disp_example} for more details.}
    \label{fig:disps1}
\end{figure*}

\begin{figure*}
    \centering
    \includegraphics[width=0.33\linewidth]{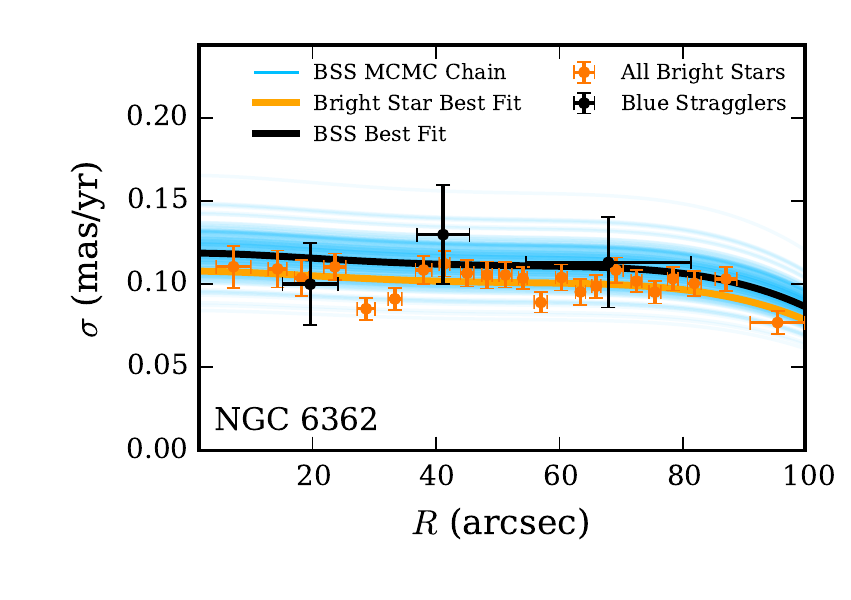}
    \includegraphics[width=0.33\linewidth]{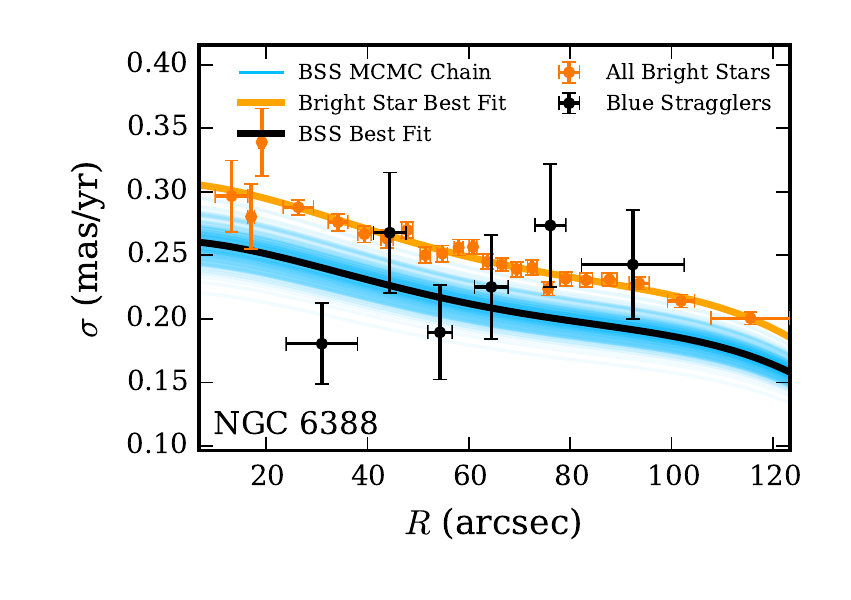}
    \includegraphics[width=0.33\linewidth]{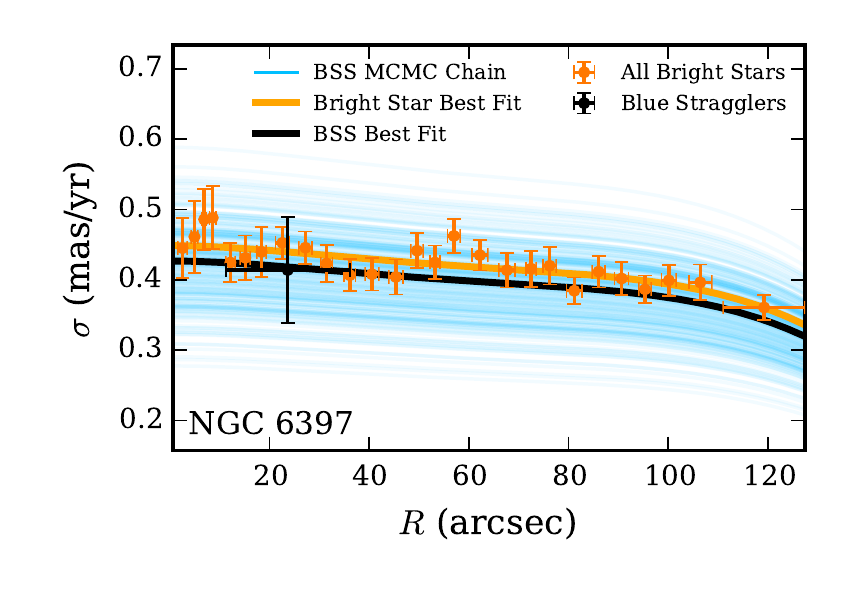}
    \includegraphics[width=0.33\linewidth]{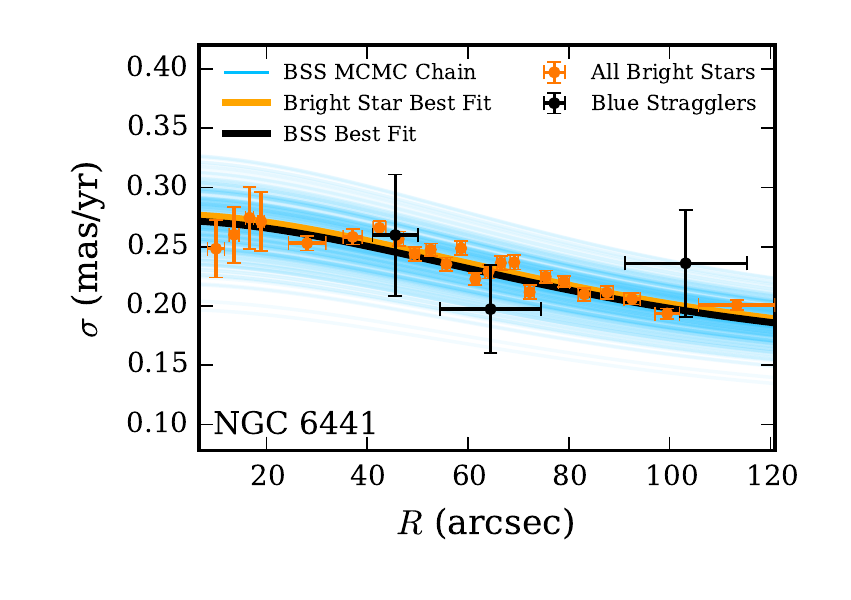}
    \includegraphics[width=0.33\linewidth]{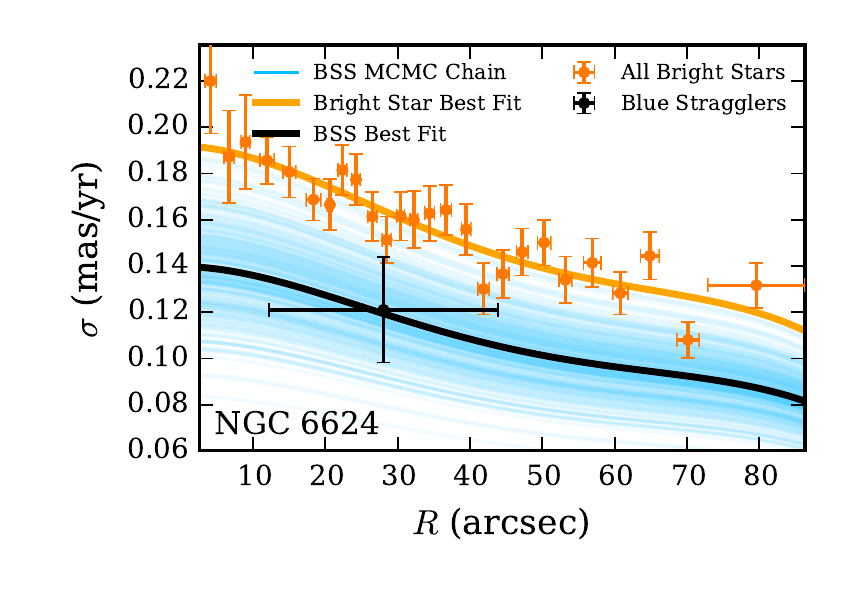}
    \includegraphics[width=0.33\linewidth]{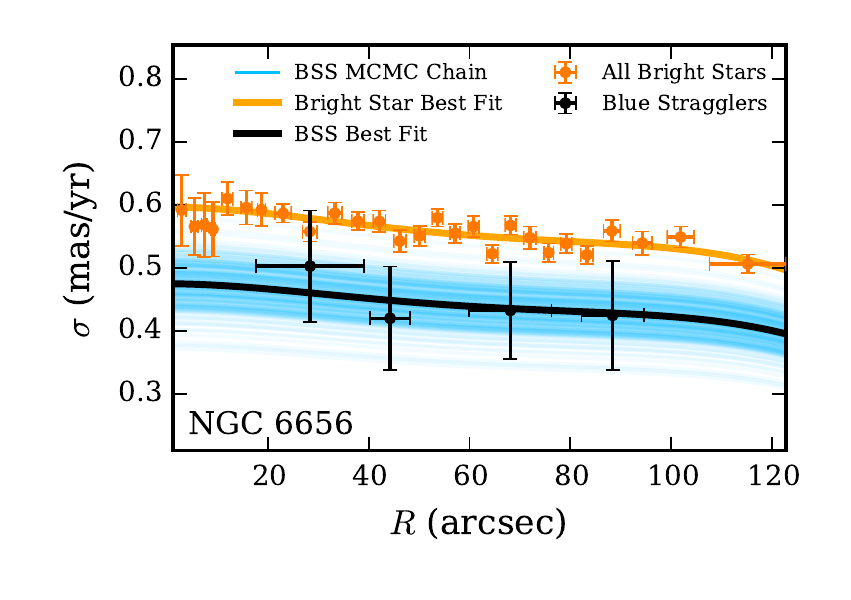}
    \includegraphics[width=0.33\linewidth]{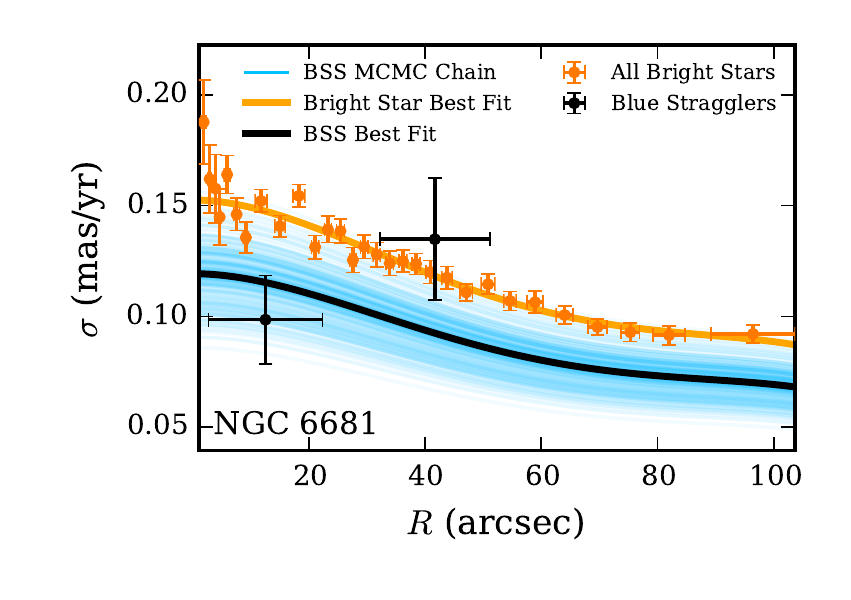}
    \includegraphics[width=0.33\linewidth]{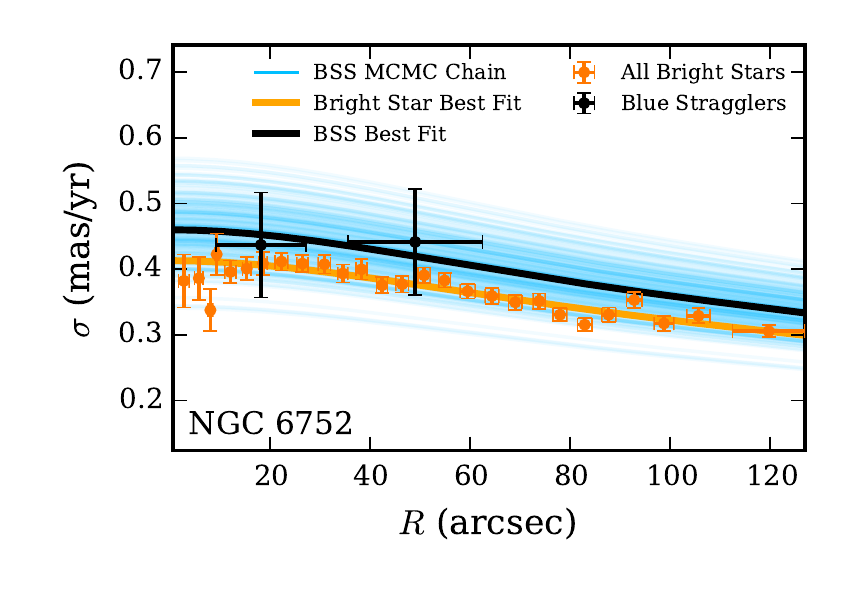}
    \includegraphics[width=0.33\linewidth]{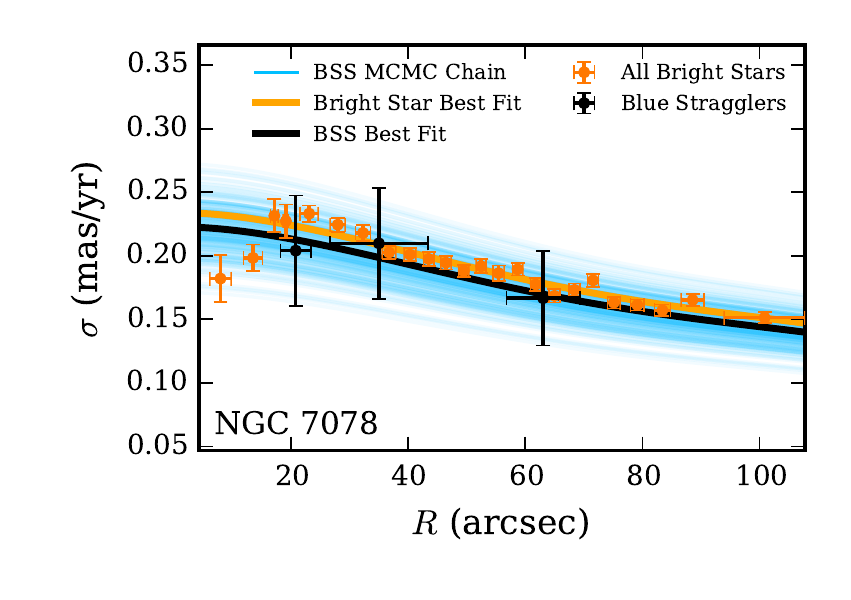}
    \caption{Velocity dispersion profiles for NGC\,6362, NGC\,6388, NGC\,6397, NGC\,6441, NGC\,6624, NGC\,6656, NGC\,6681, NGC\,6752 and NGC\,7078. See the caption to \autoref{fig:disp_example} for more details.}
    \label{fig:disps3}
\end{figure*}

We show similar velocity-dispersion profiles for all clusters in Figures~\ref{fig:disps1} and \ref{fig:disps3}. As for NGC\,362, many clusters have BSS populations that exhibit lower dispersions than the other bright stars, indicating that they are indeed more massive. However, for some clusters, the BSS dispersion profiles are similar to, or even slightly higher than, the bright-star dispersion profiles. This could indicate that these clusters are not very relaxed and so the BSSs have not had enough time to come into equipartition.

This is not unexpected. As briefly discussed earlier, the radial distributions of BSSs in GCs can be flat, bimodal, or unimodal, depending on their dynamical histories; and, in fact, these radial-distribution shapes can be used as a dynamical `clock' \citep{ferraro2012}.\footnote{Although recent results from Monte Carlo and N-body models have challenged this claim \citep{hypki2016}.} Over time, BSSs in a cluster will relax and sink towards the centre via dynamical friction; relaxation times in cluster cores are shorter than in the outer regions, so the centres will tend to relax first. Consequently, dynamically-old clusters are expected to have a centrally-concentrated population of BSSs as all of the BSSs will have had time to sink towards the centre; clusters of dynamically-intermediate age are expected to show a bimodal radial distribution because the central BSSs will have relaxed and moved to the centre, while the outer BSSs will not have had enough time to relax; and dynamically-young clusters are expected to have flat radial distributions because none of the BSSs will have had enough time to relax.

Following the same lines of reasoning, for dynamically-old clusters, we would expect to see a clear separation between the BSS dispersions and the evolved-star dispersions; for the dynamically-intermediate age clusters, we would expect to see a clear separation between the BSSs dispersions and evolved-star dispersions near the centre, but little or no difference in the dispersions in the outer regions; and, for the dynamically-young clusters, we would expect little or no difference in the dispersions across the whole cluster. Though, as our BSS populations are mostly restricted to the central regions of clusters, the dynamically-old and dynamically-intermediate age clusters will likely be largely indistinguishable in this study.

Now let us consider the clusters in our sample for which radial BSS distributions have been measured. NGC\,5139 ($\omega$~Cen) shows a flat radial distribution \citep{ferraro2006} and we find that the BSS and evolved-star dispersions are almost identical, so both result are consistent and suggest that the cluster is dynamically-young. NGC\,104 (47~Tuc) shows a bimodal radial distribution \citep{ferraro2004} and we see a clear offset in the BSS velocity dispersion profile, so again the results are consistent, but this time suggesting that the cluster is of dynamically-intermediate age. The radial distributions and kinematics are also consistent for NGC\,6388 (where both the bimodal radial distribution \citep{dalessandro2008} and the clear offset between the BSS and evolved-star kinematics suggest a dynamically-intermediate age) and NGC\,362 (where the centrally-concentrated radial distribution \citep{dalessandro2013} and the offset in the kinematics are consistent with the cluster being dynamically old). Our conclusions for these four clusters are also consistent with the number of relaxation times that we estimate the clusters to have experienced (see \autoref{table:results}).

However, for both NGC\,6752 and NGC\,5904, the two results are at odds: the radial distributions are clearly bimodal \citep{sabbi2004,lanzoni2007}, suggesting that the clusters are of dynamically-intermediate age, whereas we find that the BSS velocity dispersions suggest that they are dynamically young. NGC\,6752 is a core-collapsed cluster, so our results may indicate that there are additional dynamical processes at work in its very dense core that have washed out any velocity dispersion differences, although other known core-collapsed clusters, such as NGC\,6681, do show clear velocity dispersion differences. Further, NGC\,5904 is not thought to be core-collapsed, so the reasons for the mismatch in this case is unclear.

It is also worth noting that any systematics in our analysis could cause us to overestimate some dispersion profiles and underestimate others; these will be accounted for in our final average, but may explain why the results are inconsistent for NGC\,6752.

\begin{table}
    \caption{Blue straggler radial dispersion profiles.}
    \label{table:profiles}
    \centering
    
    \begin{tabular}{cccccccc}
        \hline
        \hline
        Cluster & $N_\mathrm{BSS}$ & R & $\Delta$R & $\sigma$ & $\Delta\sigma$\\
        & & (arcsec) & (arcsec) & (mas/yr) & (mas/yr) \\
        (1) & (2) & (3) & (4) & (5) & (6) \\
        \hline
        NGC\,104 & 8 & 30.2056 &  6.2296 & 0.5777 & 0.1048 \\
                 & 9 & 49.7092 &  4.3549 & 0.3774 & 0.0647 \\
                 & 8 & 74.4922 & 12.5776 & 0.4158 & 0.0758 \\
        NGC\,288 & 9 & 22.0483 &  8.5347 & 0.0622 & 0.0120 \\
                 & 9 & 37.7978 &  4.4031 & 0.0533 & 0.0110 \\
        \hline
    \end{tabular}
    
    \qquad
    
    \raggedright
    \textbf{Notes:} Columns: (1) cluster ID; (2) number of BSS in the bin; (3) mean radius of the bin; (4) error on the mean radius; (5) binned velocity dispersion; (6) error on the binned velocity dispersion.
    
    \qquad
    
    (This table is available in its entirety in a machine-readable form. A portion is shown here as an example of form and content.)
\end{table}

We provide our binned BSS velocity-dispersion profiles in \autoref{table:profiles}.

\subsection{Estimates of equipartition}
\label{sect:eta}

We have directly measured $\alpha$ and would like to use these measurements to determine the average mass of the BSS population via \autoref{eqn:alpha}. To do this, we must estimate $\eta$. A direct measurement would require velocity-dispersion profiles for stars well below the turnoff mass and is beyond the scope of this paper. Instead, we turn to simulations of GCs.

\citet{trenti2013} used a set of N-body simulations to show that GCs do not achieve complete energy equipartition even after several relaxation times, and that even more massive objects, such as compact remnants and BSSs, approach a value of $\eta$ well below 0.5 within the core (their Figure~1). More recently, \citet{bianchini2016b} studied a set of 7 Monte Carlo cluster simulations, with varying concentrations, binary fractions, and total cluster masses. Each simulation was analysed at 4, 7 and 11 Gyr, yielding 21 snapshots in total. They showed that the degree of equipartition reached by stars in a simulated cluster depends on stellar mass, such that stars more massive than some threshold mass $\Meq$ are in complete equipartition (where $\eta = 0.5$), while stars less massive than $\Meq$ have values of $\eta$ that vary linearly as a function of stellar mass. That is,
\begin{equation}
    \eta \left( M \right) = \begin{cases}
        \frac{1}{2} \frac{M}{\Meq} & \mbox{if } M \le \Meq, \\
        \frac{1}{2} & \mbox{if } M > \Meq.
        \end{cases}
    \label{eqn:eta}
\end{equation}
They also showed that the threshold mass $\Meq$ varied from cluster to cluster and was strongly correlated with the number of relaxation times experienced by the cluster $\nrel = \Tage / \Trc$, where $\Tage$ is the age of the cluster and $\Trc$ is the core relaxation time, such that,
\begin{equation}
    \Meq = 1.55 + 4.10 \, \nrel^{-0.85},
    \label{eqn:Meq}
\end{equation}
(see panel C of their Figure 6); this correlation is independent of concentration, binary fraction or initial mass. So it is clear that we cannot assume that the clusters have reached full equipartition, and we must consider both the approximate mass of the BSSs and the relaxation of the cluster when determining values of $\eta$.

To begin, we determine the degree of relaxation experienced by our clusters. \citet{vandenberg2013} estimated ages -- via isochrone fitting near the MSTO -- for 55 Milky Way GCs, 15 of which overlap with our sample, leaving 4 of our clusters without age estimates. We split the clusters with age estimates into two groups based on their $\feh$ metallicities \citep[taken from][2010 edition]{harris1996}, denoting clusters with $\feh \le -1.5$ as metal poor and clusters with $\feh > -1.5$ as metal rich. Next we take the average ages separately of the metal-poor and metal-rich clusters and assign the appropriate age to the remaining 4 clusters depending on their metallicity (rounded to the nearest 0.25~Gyr to match the precision of the \citet{vandenberg2013} ages). Combining these age estimates with estimates of the core relaxation times \citep[also from][2010 edition]{harris1996}, we estimate $\nrel = \Tage / \Trc$. Then we use these $\nrel$ estimates, along with \autoref{eqn:Meq}, to estimate $\Meq$ values for each cluster.

Now, let us address the issue of stellar mass. The turnoff mass is typically around 0.8~\Msun for Galactic GCs, and we expect that BSS masses will typically fall somewhere between the turnoff mass and twice the turnoff mass, so most BSSs within our sample should have masses between 0.8~\Msun and 1.6~\Msun. For our purposes, we do not need to evaluate $\eta$ as a function of $M$, instead we require an average $\eta$ across this mass range. So we use our $\Meq$ estimates and \autoref{eqn:eta} to estimate $\eta$ at $M=1.2$~\Msun (the middle of the range of interest) for each cluster, and adopt these as representative $\eta$ values across the putative BSS range.

\begin{figure}
    \centering
    \includegraphics[width=\linewidth]{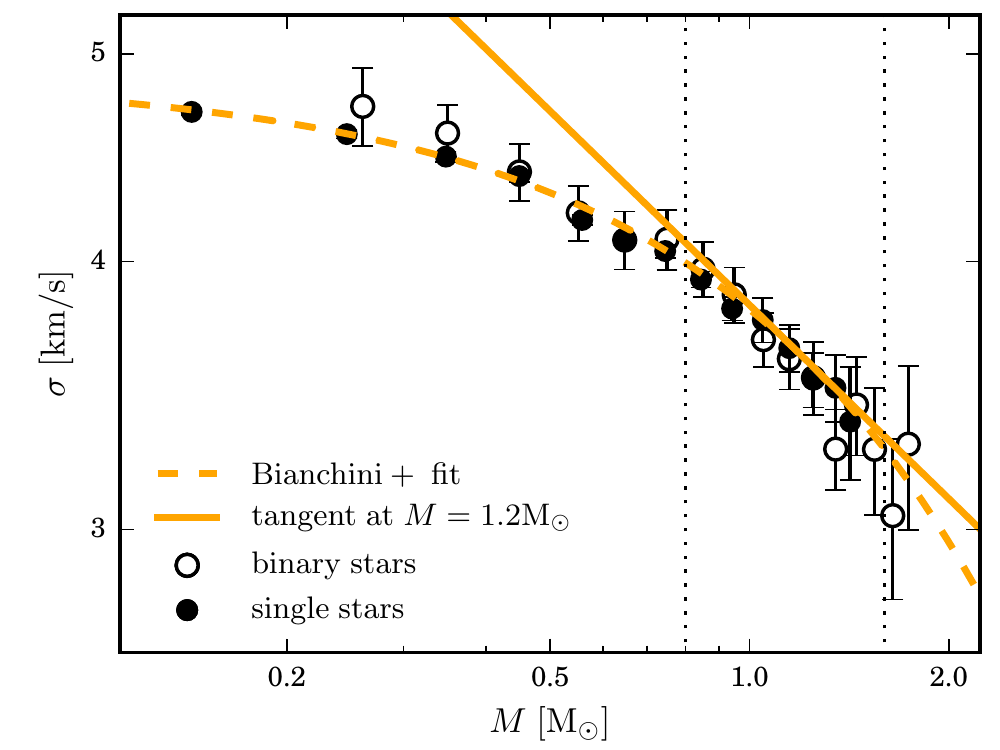}
    \caption{Velocity dispersion as a function of stellar mass for single stars (solid circles) and binary stars (open circles) as predicted by one of the Monte Carlo simulations of \citet{bianchini2016b}. The dashed orange line corresponds to an exponential fit across the full range of stellar masses; it is clear that a straight-line fit to the whole mass range would be poor. The solid orange line shows the slope of the fit at $M=1.2$~\Msun, which we use to approximate the fit across the 0.8-1.6~\Msun range (highlighted by the dotted lines) in which we expect to find BSSs; a straight line is a reasonable fit in this limited mass range, and the slope of this line provides an estimate of $\eta$, the degree of equipartition reached by the simulation.}
    \label{fig:etasims}
\end{figure}

In \autoref{fig:etasims}, we show velocity dispersion as a function of stellar mass for Simulation 1 from \citet{bianchini2016b}. The filled circles show the dispersion profile for single stars and the open circles show the dispersion profile for binary systems; these are consistent across the whole mass range. This is important as it allows us to henceforth consider the single and binary populations together. The dashed orange line shows an exponential fit to the dispersion profile from \citep[][their equation 3]{bianchini2016b}; the instantaneous slope of this line gives the value of $\eta$ for any given stellar mass (\autoref{eqn:eta}). The solid orange line shows the tangent of the exponential fit at $M=1.2$~\Msun (the middle of the putative BSS range). The dotted lines mark the 0.8-1.6~\Msun mass range in which we are interested for our BSS study.

Across the whole mass range, the exponential function clearly performs better than a simple straight-line fit. However, in the BSS mass range, we do find that the dispersion profile can be well approximated by a straight line, implying that $\eta$ can be assumed constant in this range. Further, the deviation of the exponential fit from the tangent evaluated at $M=1.2$~\Msun is small in the BSS mass range, indicating that the value of $\eta$ at $M=1.2$~\Msun can indeed by used a representative value, as we have done.

Finally, we must consider the uncertainty in the $\eta$ values we have determined. There are a number of sources of uncertainty, including: uncertainty on the ages and relaxation times used to calculate $\nrel$; scatter in \autoref{eqn:Meq}; scatter in \autoref{eqn:eta}. Furthermore, the \citet{bianchini2016b} simulations start from a specific set of initial conditions that may not accurately represent the initial conditions of our clusters, and the 11 Gyr evolution of the simulated clusters may also not accurately reproduce the history of our clusters \citep[see also][]{bianchini2016a}. Finally, the value of $\eta$ at $M=1.2$~\Msun is intended to be representative of $\eta$ across the range 0.8-1.6~\Msun, but we do know that $\eta$ does change with stellar mass, so this may be a further source of uncertainty. Also consider that our choice of $0.8-1.6$~\Msun as an expected BSS mass range was motivated by typical turnoff mass in clusters, but there will also be cluster-to-cluster variations in the turnoff mass. To encompass all of these sources, we adopt a generous systematic uncertainty of $\sfrac{\eta}{3}$ for each cluster.

We provide $\feh$ metallicities, ages, relaxation times, $\nrel$ estimates, $\Meq$ estimates and $\eta$ estimates in \autoref{table:results}.

\begin{table*}
    \caption{Properties and results for Galactic globular clusters.}
    \label{table:results}
    \centering
    
    \begin{tabular}{cccccccccccc}
        \hline
        \hline
        Cluster ID & $\feh$ & $\Tage$ & $r_{\rm c}$ & $\log_{10} \Trc$ & $\log_{10} \nrel$ & $\Meq$ & $\eta$ & $\Mto$ & $\alpha$ & $f$ & $\Mbss$ \\
        & (dex) & (Gyr) & (arcsec) & (Gyr) &  & (\Msun) &  & (\Msun) &  &  & (\Msun) \\
        (1) & (2) & (3) & (4) & (5) & (6) & (7) & (8) & (9) & (10) & (11) & (12) \\
        \hline \\[-1em]
        NGC\,104  & -0.72 & 11.75 &  21.6 & -1.16 & 2.23 & 1.60 & 0.37 & 0.87 & $0.77 \pm 0.06$ & $2.04_{-0.41}^{+0.64}$ & $1.78_{-0.35}^{+0.56}$ \\[0.4em]
        NGC\,288  & -1.32 & 11.50 &  81.0 & -0.01 & 1.07 & 2.05 & 0.29 & 0.79 & $0.80 \pm 0.07$ & $2.18_{-0.55}^{+0.88}$ & $1.71_{-0.43}^{+0.69}$ \\[0.4em]
        NGC\,362  & -1.26 & 10.75 &  10.8 & -1.24 & 2.27 & 1.60 & 0.38 & 0.80 & $0.69 \pm 0.04$ & $2.72_{-0.52}^{+0.91}$ & $2.18_{-0.42}^{+0.73}$ \\[0.4em]
        NGC\,1851 & -1.18 & 11.00 &   5.4 & -1.57 & 2.61 & 1.57 & 0.38 & 0.83 & $0.85 \pm 0.07$ & $1.53_{-0.27}^{+0.45}$ & $1.27_{-0.22}^{+0.38}$ \\[0.4em]
        NGC\,2808 & -1.14 & 11.00 &  15.0 & -0.76 & 1.80 & 1.67 & 0.36 & 0.83 & $0.95 \pm 0.05$ & $1.15_{-0.14}^{+0.20}$ & $0.95_{-0.11}^{+0.17}$ \\[0.4em]
        NGC\,5139 & -1.53 & 12.75 & 142.2 &  0.60 & 0.51 & 3.07 & 0.20 & 0.76 & $1.02 \pm 0.04$ & $0.89_{-0.15}^{+0.23}$ & $0.68_{-0.12}^{+0.17}$ \\[0.4em]
        NGC\,5904 & -1.29 & 11.50 &  26.4 & -0.72 & 1.78 & 1.68 & 0.36 & 0.80 & $0.99 \pm 0.04$ & $1.03_{-0.23}^{+0.36}$ & $0.82_{-0.18}^{+0.29}$ \\[0.4em]
        NGC\,5927 & -0.49 & 10.75 &  25.2 & -0.61 & 1.64 & 1.72 & 0.35 & 0.89 & $0.87 \pm 0.04$ & $1.50_{-0.19}^{+0.27}$ & $1.33_{-0.17}^{+0.25}$ \\[0.4em]
        NGC\,6266 & -1.18 & 11.25 &  13.2 & -1.10 & 2.15 & 1.61 & 0.37 & 0.81 & $1.10 \pm 0.09$ & $0.77_{-0.17}^{+0.22}$ & $0.62_{-0.13}^{+0.18}$ \\[0.4em]
        NGC\,6341 & -2.31 & 12.75 &  15.6 & -1.04 & 2.15 & 1.61 & 0.37 & 0.76 & $0.78 \pm 0.05$ & $1.96_{-0.31}^{+0.51}$ & $1.49_{-0.23}^{+0.39}$ \\[0.4em]
        NGC\,6362 & -0.99 & 12.50 &  67.8 & -0.20 & 1.30 & 1.87 & 0.32 & 0.81 & $1.10 \pm 0.10$ & $0.74_{-0.19}^{+0.28}$ & $0.60_{-0.15}^{+0.22}$ \\[0.4em]
        NGC\,6388 & -0.55 & 11.25 &   7.2 & -1.28 & 2.33 & 1.59 & 0.38 & 0.85 & $0.85 \pm 0.04$ & $1.53_{-0.18}^{+0.27}$ & $1.31_{-0.16}^{+0.22}$ \\[0.4em]
        NGC\,6397 & -2.02 & 13.00 &   3.0 & -4.06 & 5.17 & 1.55 & 0.39 & 0.76 & $0.95 \pm 0.12$ & $1.12_{-0.27}^{+0.54}$ & $0.84_{-0.20}^{+0.42}$ \\[0.4em]
        NGC\,6441 & -0.46 & 11.25 &   7.8 & -1.07 & 2.12 & 1.61 & 0.37 & 0.84 & $0.98 \pm 0.08$ & $1.05_{-0.21}^{+0.29}$ & $0.88_{-0.17}^{+0.25}$ \\[0.4em]
        NGC\,6624 & -0.44 & 11.25 &   3.6 & -2.39 & 3.44 & 1.55 & 0.39 & 0.87 & $0.73 \pm 0.10$ & $2.28_{-0.67}^{+1.26}$ & $1.99_{-0.58}^{+1.09}$ \\[0.4em]
        NGC\,6656 & -1.70 & 12.50 &  79.8 & -0.47 & 1.57 & 1.74 & 0.34 & 0.77 & $0.80 \pm 0.06$ & $1.94_{-0.37}^{+0.61}$ & $1.49_{-0.28}^{+0.47}$ \\[0.4em]
        NGC\,6681 & -1.62 & 12.75 &   1.8 & -3.18 & 4.29 & 1.55 & 0.39 & 0.76 & $0.78 \pm 0.08$ & $1.86_{-0.41}^{+0.76}$ & $1.42_{-0.31}^{+0.58}$ \\[0.4em]
        NGC\,6752 & -1.54 & 12.50 &  10.2 & -2.12 & 3.22 & 1.56 & 0.39 & 0.77 & $1.11 \pm 0.10$ & $0.76_{-0.17}^{+0.22}$ & $0.59_{-0.13}^{+0.17}$ \\[0.4em]
        NGC\,7078 & -2.37 & 12.75 &   8.4 & -1.16 & 2.27 & 1.60 & 0.38 & 0.76 & $0.95 \pm 0.08$ & $1.15_{-0.23}^{+0.30}$ & $0.87_{-0.17}^{+0.24}$ \\[0.4em]
        \hline
        Mean & \dots & \dots & \dots & \dots & \dots & \dots & \dots & \dots & $0.90 \pm 0.03$ & $1.48 \pm 0.13$ & $1.20 \pm 0.11$ \\[0.4em]
        \hline
    \end{tabular}
    
    \qquad
    
    \raggedright
    \textbf{Notes.} Columns: (1) NGC identification; (2) $\feh$ metallicity from \citet[][2010 edition]{harris1996}; (3) cluster age (in Gyr), mostly taken from \citet{vandenberg2013}, others derived from average \citet{vandenberg2013} values and $\feh$ metallicity -- see text for details; (4) core radius $r_{\rm c}$ (in arcsec) \citep[][2010 edition]{harris1996}; (5) logarithm of the core relaxation time (in Gyr) \citep[][2010 edition]{harris1996}; (6) logarithm of the number of relaxation times $\nrel = \Tage/\Trc$; (7) equipartition mass $\Meq$ (in \Msun) above which stars are in complete energy equipartition \citep[see][for further details]{bianchini2016b}; (8) equipartition parameter $\eta$; (9) MSTO mass $\Mto$ (in \Msun); (10) $\alpha = (\Mbss/\Mto)^{-\eta}$ estimated from our dispersion profiles; (11) average BSS mass as a multiple of the MSTO mass, $f = \Mbss/\Mto$; (12) average BSS mass $\Mbss$ (in \Msun).
    
    \qquad
    
    (This table is available in machine-readable form.)
\end{table*}

\subsection{Blue Straggler Mass Fractions}
\label{sect:massfracs}

Now, we are ready to estimate the mass ratio for each cluster. We begin with the values of $\alpha$ returned by the MCMC sampling in \autoref{sect:bssdisps} that we will turn into a mass ratio
\begin{equation}
    f \equiv \frac{\Mbss}{\Mto},
\end{equation}
by solving \autoref{eqn:alpha} for
\begin{equation}
    f = \alpha^{-\frac{1}{\eta}}.
    \label{eqn:f}
\end{equation}

To accurately propagate our uncertainty in $\eta$, we draw 1000 values of $\eta$ from a boxcar distribution with half-width $\sfrac{\eta}{3}$ centered on our best estimate of $\eta$ for each cluster (\autoref{table:results}). We then take the median of all mass fractions returned by this method to be our best estimate of the average BSS mass ratio in the cluster, and the distance between the 16th and 84th percentiles to be the lower and upper error bars, respectively. Mass ratio estimates for each cluster are presented in \autoref{table:results}.

The reduced $\chi^2$ for the sample compared to its unweighted mean is 3.4. That this value is larger than unity indicates that there is more cluster-to-cluster scatter in our inferred BSS masses than can be explained by our random errors. That is, there are likely further sources of systematic uncertainty, at a level comparable to the random errors, for which we have not accounted.

For this reason, it is not appropriate to include the error bars when calculating a sample mean for all the clusters, so we use a simple unweighted average to calculate the mean sample BSS mass ratio. Similarly, to estimate the error on the mean we use $\sigma/\sqrt{N}$, where $\sigma$ is the scatter between measurements for different clusters. This yields a result that is based only on the scatter between the mass ratios for each cluster, and ignores the random error bars in the individual measurements.

Calculated in this way, we find an average mean mass ratio for the sample of $1.50 \pm 0.14$. The error bar on our final mean result is symmetric; this is reasonable as the distribution of BSS masses inferred for different clusters is not strongly asymmetric (unlike the random errors for individual clusters, which do often tend to be strongly asymmetric). We have experimented with other statistics for calculating the sample mean, and find that the results of alternative methods are generally consistent with this result within the error bars.

\begin{figure}
    \centering
    \includegraphics[width=0.99\linewidth]{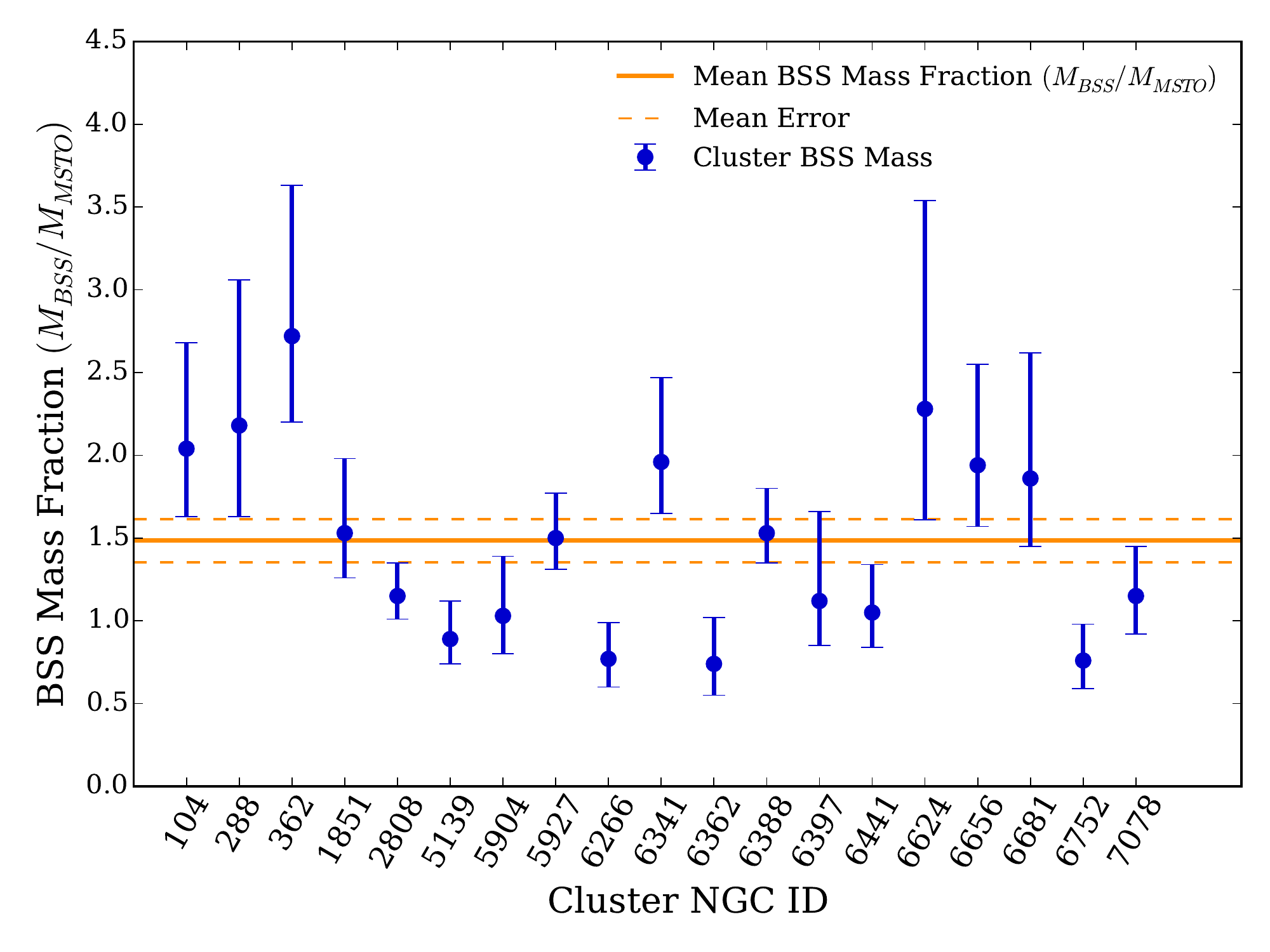}
    \caption{Estimates of BSS mass as a multiple of MSTO mass for each cluster. The orange line shows the mean BSS mass ratio for the sample and the orange dashed lines represent the standard error on the mean.}
    \label{fig:fmasses}
\end{figure}

\autoref{fig:fmasses} shows mass-ratio estimates for all clusters in our sample. The orange line represents the average mass ratio across all the clusters, and the dashed lines represent the standard error on the mean.

\subsection{Blue Straggler Masses}

So far, we have estimated BSS masses as a multiple of the turnoff mass in a cluster. To determine the intrinsic mass of BSSs, we require estimates for the turnoff masses, which we obtain via isochrone fitting using isochrones from the Dartmouth Stellar Evolution Database \citep{dotter2008}. Interpolating between isochrones is beyond the scope of this paper, so to select representative isochrones for each cluster, we require: $\feh$ metallicities to the nearest 0.5~dex; $\alpha$-element abundances $\afe$ to the nearest 0.2~dex; and ages to the nearest 0.5~Gyr.

As we did in \autoref{sect:eta}, we use $\feh$ values from \citet[][2010 edition]{harris1996} and ages from \citet{vandenberg2013} (see also \autoref{table:results}), appropriately rounded. In general, GCs have $\alpha$-element abundances $\afe \sim 0.3$ \citep[eg.][]{carney1996}, but they do show a mild correlation with $\feh$ metallicity, such that more metal-poor clusters are more $\alpha$ enhanced \citep[eg.][]{kirby2008,johnson2011}. As such, for metal-poor clusters with $\feh \le -1.5$, we assume $\alpha$-element abundance $\afe = 0.4$; for metal-rich clusters with $\feh \geq -1.5$, we assume $\alpha$-element abundance $\afe = 0.2$.

We use these metallicities, $\alpha$-element abundances and ages to extract representative isochrones and then we adjust the isochrone magnitudes for distance and extinction using distances and reddening values from \citet[][2010 edition]{harris1996} and extinction coefficients from \citet{sirianni2005}. We interpolate along the isochrone to extract mass estimates for our stars based on their apparent magnitudes. These mass estimates may not be reliable for stars that have evolved off the main-sequence. However, this is of no consequence here as we are only interested in the masses of stars near the turnoff for which this method is robust. Finally, we adopt the median mass of all stars within 0.05~mag of the MSTO as the turnoff mass of the cluster for all clusters, except NGC\,6266 for which we have no isochrone fit due to its unusual combination of filters; instead, for NGC\,6266 we adopt a value of 0.81~\Msun, which is the average turnoff mass for all of the other clusters in our sample.

The actual average mass estimate for BSSs within a cluster is then simply the mass ratio derived in \autoref{sect:massfracs} multiplied by our estimate of the MSTO mass; uncertainties in the turnoff mass are negligible compared to the uncertainty in the mass ratio. The BSS mass estimates and our turnoff mass estimates are also given in \autoref{table:results}.

As discussed in detail in \autoref{sect:massfracs}, we calculate an unweighted mean and the standard error on the mean to estimate an average BSS mass of $\Mbss = 1.22 \pm 0.12$~\Msun for the whole sample.

\section{Discussion}
\label{sect:discussion}

Our best mass estimate of $\Mbss = 1.22 \pm 0.12$~\Msun is in very good agreement with the average mass of $\Mbss = 1.22 \pm 0.06$~\Msun taken from 35 individual BSSs in the literature \citep{shara1997, gilliland1998, demarco2005, fiorentino2014}.\footnote{We include the variable BSSs from \citet{demarco2005} in this average, but neglect the BSSs from \citet{fiorentino2014} for which the pulsation mode was ambiguous.}

As we have derived our BSS masses from cluster dynamics, we have measured the average \textit{total} mass of the BSS systems, which includes any possible binary companions. The BSS population of open cluster NGC\,188 has a binary fraction of 76\% \citep{mathieu2009}; GCs are typically more dense environments than open clusters, which may affect the fraction of BSSs found in binaries, however many common formation channels for BSSs involve binaries in some fashion so the binary fraction is likely to be high in GCs as well. Furthermore, close 3-body interactions involving binary systems tend to eject the \textit{least} massive object, and so, as some of the most massive objects in a cluster, BSSs that exist in a binary system are likely to remain a part of one.

There are three main formation theories for blue stragglers: stable mass-transfer in binary systems, stellar mergers, and stellar collisions; although cluster dynamics are complicated so BSS formation histories seldom follow just one of these channels \citep[eg.][]{chatterjee2013}. Indeed, \citet{leigh2016} recently showed that binary mass-transfer can be interrupted by a dynamical encounter with another star, particularly in lower-mass clusters. Nevertheless, let us consider each mechanism in turn and consider the resulting BSS.

In the case of stable mass transfer in a binary system, the more-massive star in a binary fills its Roche lobe and transfers mass to its companion \citep{mccrea1964}. We would expect BSSs formed via mass transfer to retain at least a helium white dwarf companion with a mass on the order of 0.5~\Msun.\footnote{This represents the compact remnant of an evolved donor star after mass transfer has ceased \citep[see][]{gosnell2014}.} If the mass-transfer formation channel for BSSs is active in GCs, then we might expect our dynamical mass estimates to be somewhat larger than those derived from spectra or pulsations which only consider the BSS itself and not the additional mass of a companion.

BSSs resulting from mergers are expected to form via two alternative pathways: 1) unstable mass transfer in a binary system can lead to the complete merger of the two stars \citep{chen2009}, leaving behind a single BSS; or 2) the Kozai effect \citep{kozai1962} can cause the inner two stars of a hierarchical triple to merge \citep{perets2009}, leaving behind a BSS with the third star as a binary companion. The second formation mechanism is thought to be significant in open clusters, but not in the more dense environments found in GCs \citep{perets2009}. So the primary merger channel active in GCs is likely to be unstable mass transfer; in this case, we would expect our dynamical estimates to be consistent with those derived from spectra or pulsations.

Stellar collisions can occur as single-single encounters, binary-single encounters, or binary-binary encounters \citep[eg.][]{hut1983,sigurdsson1993}, or even triple-single, triple-binary or triple-triple encounters \citep[eg.][]{leigh2011a}. It is likely that single-single encounters may result in a single BSS, however binary-single, binary-binary and other multiple interactions are likely to leave a BSS that exists as part of a binary or multiple system. As binary or multiple encounters are generally more common than single-single encounters in GCs \citep[eg.][]{leonard1989,leigh2011b}, collisions are more likely to result in a BSS with a binary companion. Again, in this case, we would expect our dynamical BSS mass estimates to be higher than the literature values for individual BSSs.

Simulations, such as those studied in \citet{chatterjee2013}, suggest that collisions and stable-mass transfer are the dominant mechanisms for BSS formation. That our dynamical estimates are in such good agreement with previous studies may imply that the binary fraction of BSSs is lower than expected, and may further imply that stellar mergers resulting from unstable mass transfer play a more significant role in BSS formation in GCs than predicted.

However, we must consider that the individual BSS measurements from previous studies may not represent an unbiased sample of the BSS population: \citet{demarco2005} only included stars with effective temperatures greater than 5750\,K whereas \citet{gilliland1998} and \citet{fiorentino2014} focused exclusively on pulsating BSSs. By contrast, we have measured the average mass for \textit{all} 598 BSSs detected within our sample of 19 Galactic GCs, which should provide an unbiased representation of the BSS population.

Recently, \citet{xin2015} simulated a population of BSSs formed via mass transfer to mimic the population in NGC\,7099 (M30). Within the simulation, systems that eventually form BSSs had a total mean binary mass of $1.21 \pm 0.03$~\Msun. Unfortunately, NGC\,7099 is one of the clusters for which we have insufficient BSSs in the bright-star catalog to produce a BSS dispersion profile and, thus, estimate a mass. However, we can predict a BSS mass from our data by combining our mean BSS mass fraction ($\Mbss/\Mto = 1.50 \pm 0.14$) with our turnoff mass estimate for NGC\,7099 ($\Mto = 0.76$~\Msun) to predict an average BSS mass of $\Mbss = 1.14 \pm 0.10$~\Msun, which is consistent with the \citet{xin2015} prediction to within 1$\sigma$.

Our results should be taken with a few crucial caveats:

\begin{itemize}
    \item The velocity dispersion profile of BSSs need not be a simply-scaled version of the bright-star profile, since $\eta$ has been shown to vary as a function of radius \citep[see][]{trenti2013}.
    \item The relationship between $\eta$ and cluster relaxation that arises in the simulations from \citet{bianchini2016b} may not be correct for real GCs if they evolved from initial conditions that do not exactly match the simulated clusters.
    \item All of the clusters in our sample are known to host multiple populations of stars \citep[see e.g.][]{piotto2015}. Second-generation stars typically make up a sizeable fraction of the cluster as a whole; they tend to be He-enhanced and, hence, have a lower MSTO mass, so our BSS masses may be overestimated.
\end{itemize}

As a final approach, we can define a model-independent \textit{minimum} mass ratio by assuming that each cluster is in complete energy equipartition with $\eta = 0.5$. Doing so returns a minimum average mass ratio of $f \ge 1.32 \pm 0.08$. This implies with 4$\sigma$ confidence that BSSs, on average, have masses greater than the turnoff mass.

Finally, we note that the final error bars on our mass-ratio and mass estimates include any systematic errors that are random between different clusters, but do not include the possible impact of any potential systematic errors (i.e., a bias) that would shift the mass estimates for different clusters in the same direction. We have discussed various potential sources of systematic error in our data-model comparisons, and have not identified any individual source that we expect to introduce a significant bias, but this does not prove that biases may not exist. As noted, our final estimate agrees with literature estimates based on other methods to within the random error of our measurement. This suggests than any potential systematic biases in our final estimate are no larger than the random error.

\section{Conclusions}
\label{sect:conclusions}

We have produced velocity-dispersion profiles for the BSS populations in 19 Galactic GCs based on the \textit{HST} proper-motion catalogs presented in \citetalias{bellini2014}. From these profiles:

\begin{itemize}
    \item We found that BSSs typically have lower velocity dispersions than stars at the MSTO, as one would expect for a more massive population of stars in a system with some degree of energy equipartition.
    \item We derived an average mass ratio of $\Mbss/\Mto = 1.50 \pm 0.14$ for all 598 BSSs across all 19 clusters; this corresponds to an average mass of $\Mbss = 1.22 \pm 0.12$~\Msun.
    \item We confirmed at the $4\sigma$ level that BSSs are on average more massive than the turnoff mass.
\end{itemize}

Our dynamical estimates are in very good agreement with previous estimates for BSS masses \citep{shara1997, gilliland1998, demarco2005, fiorentino2014} based on the properties of individual stars.

\qquad

We wish to thank Yu Xin for providing us with us data on mass transfer BSS simulations for M30, and Nathan Leigh for interesting discussions about blue stragglers and useful comments on the draft. We also wish to thank the anonymous referee for the useful reports that improved the discussion of our results. Support for this work was provided by grants for \textit{HST} programs AR-12845 (PI: Bellini) and AR-12648 (PI: van~der~Marel), provided by the Space Telescope Science Institute, which is operated by AURA, Inc., under NASA contract NAS 5-26555. PB acknowledges support from the International Max Planck Research School in Astronomy and Cosmic Physics at the University of Heidelberg (IMPRS-HD).

This research made use of Astropy\footnote{\url{http://www.astropy.org}}, a community-developed core Python package for Astronomy \citep{astropy2013}. This research has made use of NASA's Astrophysics Data System Bibliographic Services.

This project is part of the HSTPROMO collaboration\footnote{\url{http://www.stsci.edu/~marel/hstpromo.html}}, a set of HST projects aimed at improving our dynamical understanding of stars, clusters and galaxies in the nearby Universe through measurement and interpretation of proper motions.

\bibliographystyle{apj}
\bibliography{refs}

\end{document}